\documentclass{aa}  
\usepackage{graphicx}
\usepackage{natbib}

\bibpunct{(}{)}{;}{a}{}{,}   
\usepackage{txfonts}

\begin{document}
\title{ATCA 3\,mm observations of NGC6334I and I(N): dense cores,
  outflows and an UCH{\sc ii} region}


   \author{H.~Beuther\inst{1},
          A.J. Walsh\inst{2},
          S. Thorwirth\inst{3},
          Q. Zhang\inst{4},
          T.R. Hunter\inst{5},
          S.T. Megeath\inst{6}
          and
          K.M. Menten\inst{3}
          }

   \offprints{H.~Beuther}

   \institute{Max-Planck-Institute for Astronomy, K\"onigstuhl 17, 
              69117 Heidelberg, Germany\\
              \email{beuther@mpia.de}
         \and
              Centre for Astronomy, James Cook University,
              Townsville, QLD 4811 Australia \\
              \email{Andrew.Walsh@jcu.edu.au}
         \and
              Max-Planck-Institute for Radioastronomy, Auf dem H\"ugel 69, 
              53121 Bonn, Germany\\
              \email{sthorwirth@mpifr-bonn.mpg.de, kmenten@mpifr-bonn.mpg.de}
         \and
              Harvard-Smithsonian Center for Astrophysics, 60 Garden Street,
              Cambridge, MA 02138, USA\\
             \email{qzhang@cfa.harvard.edu}
         \and
              NRAO, 520 Edgemont Rd,
              Charlottesville, VA 22903\\
              \email{thunter@nrao.edu}
         \and
              Ritter Observatory, Department of Physics and Astronomy, 
              University of Toledo, Toledo, OH 43606-3390, USA\\
              \email{megeath@astro1.panet.utoledo.edu}
             }
\authorrunning{Beuther et al.}
\titlerunning{ATCA 3\,mm observations of NGC6334I and I(N)}

   \date{}

  \abstract
    {}
    {Investigation of the dense gas, the outflows and the continuum
      emission from the massive twin cores NGC6334I and I(N) at high
      spatial resolution.}
    {We imaged the region with the Australia Telescope Compact Array
      (ATCA) at 3.4\,mm wavelength in continuum as well as
      CH$_3$CN$(5_K-4_K)$ and HCN(1--0) spectral line emission.}
    {While the continuum emission in NGC6334I mainly traces the
      UCH{\sc ii} region, toward NGC6334I(N) we detect line emission
      from four of the previously identified dust continuum
      condensations that are of protostellar or pre-stellar nature.
      The CH$_3$CN$(5_K-4_K)$ lines are detected in all $K$-components
      up to energies of 128\,K above ground toward two protostellar
      condensations in both regions. We find line-width increasing
      with increasing $K$ for all sources, which indicates a higher
      degree of internal motions of the hotter gas probed by these
      high K-transitions. Toward the main mm and CH$_3$CN source in
      NGC6334I we identify a velocity gradient approximately
      perpendicular to the large-scale molecular outflow. This may be
      interpreted as a signature of an accretion disk, although other
      scenarios, e.g., an unresolved double source, could produce a
      similar signature as well. No comparable signature is found
      toward any of the other sources.  HCN does not trace the dense
      gas well in this region but it is dominated by the molecular
      outflows. While the outflow in NGC6334I exhibits a normal
      Hubble-law like velocity structure, the data are consistent with
      a precessing outflow close to the plane of the sky for
      NGC6334I(N). Furthermore, we observe a wide
      ($\sim$15.4\,km\,s$^{-1}$) HCN absorption line, much broader
      than the previously observed CH$_3$OH and NH$_3$ absorption
      lines. Several explanations for the difference are discussed.}
    {}
   

   \maketitle
%

\section{Introduction}

The massive twin cores NGC6334I and I(N) at a distance of 1.7\,kpc in
the southern hemisphere \citep{neckel1978,straw1989} have been
subjected to investigations for more than two decades. The two regions
are located at the north-eastern end of the much larger molecular
cloud/H{\sc ii} region complex NGC6334 (e.g.,
\citealt{rodriguez1982,gezari1982,depree1995,kraemer1999b,sandell2000,carral2002}).
The reason why NGC6334I (synonymous with NGC6334F) and NGC6334I(N) are
so interesting from a comparison point of view is that they are only
separated by approximately 1\,parsec, hence they share a similar
large-scale molecular environment, but they exhibit extremely
different characteristics likely because they are at different
evolutionary stages.

Both regions have been studied in much detail over the last decades;
recent summaries of the past observations can be found, e.g., in
\citet{hunter2006}, \citet{beuther2007b} or \citet{rodriguez2007}.
Here we just outline their main characteristics. NGC6334I is a
prototypical hot molecular core right at the head of a cometary
ultracompact H{\sc ii} (UCH{\sc ii}) region
\citep{depree1995,kraemer1995}.  It exhibits rich spectral line
emission \citep{mccutcheon2000,thorwirth2003,schilke2006}, a bipolar
outflow \citep{bachiller1990,leurini2006} and H$_2$O, OH, CH$_3$OH
class {\sc ii} and NH$_3$(3,3)/(6,6)/(8,6)/(11,9) maser emission
\citep{moran1980,forster1989,gaume1987,brooks2001,norris1993,caswell1997,walsh1998,beuther2007b,walsh2007}.
In contrast to that, up to very recently NGC6334I(N) was considered a
typical cold core since no mid-infrared and only faint near-infrared
emission was detected \citep{gezari1982,tapia1996,persi2005}.
Furthermore, weak cm continuum and class {\sc i} and {\sc ii} CH$_3$OH
maser emission was reported
\citep{carral2002,kogan1998,caswell1997,walsh1998}. The spectral line
forest is considerably less dense compared to NGC6334I
\citep{thorwirth2003}, however, a few species are stronger toward
NGC6334I(N) (\citealt{sollins2004c}, Walsh et al.~in prep.). In
addition, \citet{megeath1999} report the detection of a molecular
outflow in this region as well. In summary, both regions show signs of
active star formation, however, the southern region NGC6334I appears
to be in a more advanced evolutionary stage than the northern region
NGC6334I(N).

To better characterize this intriguing pair of massive star-forming
regions, we started a concerted campaign from cm to mm wavelengths
with the Australia Telescope Compact Array (ATCA), the Submillimeter
Array (SMA) and the Mopra single-dish telescope.  The previous ATCA
NH$_3$(1,1) to (6,6) line observations revealed compact warm gas
emission from both regions \citep{beuther2005e,beuther2007b}, and
temperatures estimated to exceed 100\,K. While toward NGC6334I(N) the
low energy NH$_3$ lines showed only extended emission, the high energy
lines finally revealed compact gas components. The NH$_3$(6,6) line
profile from NGC6334I(N) allowed speculation about a potential
accretion disk. CH$_3$OH was strong in absorption toward the southern
UCH{\sc ii} region in NGC6334I, indicative of expanding gas.  In the
mm continuum emission, \citet{hunter2006} used the SMA to resolve
several mm continuum sources toward both regions (4 in NGC6334I and 7
in NGC6334I(N)).  Furthermore, Hunter et al.~(in prep.) identified an
additional SiO outflow in NGC6334I(N) that has its orientation in
north-east south-west direction, approximately perpendicular to the
one previously reported by \citet{megeath1999}.

Here we present 3.4\,mm continuum and HCN/CH$_3$CN spectral line
observations obtained with the new 3\,mm facility at the ATCA. These
observations shed light on the outflow and dense gas properties of
both regions as well as on the continuum emission from the embedded
protostellar objects and the UCH{\sc ii} region.

\section{Observations}

\begin{table}[htb]
\caption{Line parameters}
\begin{tabular}{lrrr}
\hline \hline
Line & Freq. & $E_{\rm{u}}/k$ & $n_{\rm{crit}}^a$ \\
     & [GHz] & [K]          & [$10^5$cm$^{-3}$]\\
\hline
HCN(1--0), F=1-1    & 88.630 &   4 & 32 \\
HCN(1--0), F=2-1    & 88.632 &   4 & 32 \\
HCN(1--0), F=0-1    & 88.634 &   4 & 32 \\
CH$_3$CN$(5_4-4_4)$ & 91.959 & 128 & 7\\ 
CH$_3$CN$(5_3-4_3)$ & 91.971 &  78 & 7 \\ 
CH$_3$CN$(5_2-4_2)$ & 91.980 &  42 & 7 \\ 
CH$_3$CN$(5_1-4_1)$ & 91.985 &  20 & 6 \\ 
CH$_3$CN$(5_0-4_0)$ & 91.987 &  13 & 6 \\ 
\hline \hline
\end{tabular}
~\\ {\footnotesize For all observed lines the columns list (left to right) the species/quantum numbers, frequencies, upper level energy states $E_{\rm{u}}/k$ ($k$ is the Boltzmann constant) and critical densities $n_{\rm{crit}}$.}\\
{\footnotesize $^a$ The critical densities $n_{\rm{crit}}=A/\gamma$ are calculated at 60\,K (Einstein coefficient $A$ and collisional rate $\gamma$). For HCN, $A$ and $\gamma$ are taken from LAMBDA \citep{schoeier2005}. For CH$_3$CN, $A$ is calculated from $A=0.3\lambda_{100}^{-3}\mu ^2$ ($\lambda_{100}$ in units of 
  100\,$\mu$m and $\mu=3.9$\,debye), and the corresponding $\gamma$-values are from \citet{pei1995a,pei1995b}. For a line optical depth $\tau >1$, $n_{\rm{crit}}$ has to be multiplied by $1/\tau$ \citep{tielens2005}.}
\label{lines}
\end{table}

The two regions NGC6334I \& I(N) were observed in May 2006 during two
nights with the ATCA in the H214 configuration that results in
projected baselines between 14 and 73\,k$\lambda$ at 88\,GHz. The
weather conditions at Narrabri were excellent with approximate
precipitable water vapor of $\sim$10\,mm and measured system
temperatures between 150 and 670\,K. The phase reference centers were
R.A.  (J2000) $17^h20^m53^s.44$, Decl.~(J2000) $-35^{\circ}47'02''.2$
for NGC6334I and R.A. (J2000) $17^h20^m54^s.63$, Decl.~(J2000)
$-35^{\circ}45'08''.9$ for NGC6334I(N). The velocities relative to the
local standard of rest ($v_{\rm{lsr}}$) for NGC6334I and NGC6334I(N)
are $\sim -7.6$ and $\sim -3.3$\,km\,s$^{-1}$, respectively. We
observed the 3.4\,mm continuum emission at 88.4\,GHz with a bandwidth
and spectral resolution of 128 and 1\,MHz, respectively. The spectral
range for the continuum was checked to be line-free based on previous
MOPRA single-dish observations of that region (Walsh et al., in
prep.). During one night, simultaneously with the continuum emission,
we observed the HCN(1--0) line at 88.632\,GHz. In the second night,
the CH$_3$CN$(5_K-4_K)$ transitions at $\sim$91.98\,GHz were targeted
averaging two polarizations to achieve better signal-to-noise ratio.
For more details on the spectral lines, see Table \ref{lines}.  A good
uv-coverage was obtained through regular switching between both
sources and the gain calibrators 1742-289 and 1759-39. The channel
separation of the HCN observations was 0.25\,MHz
($\sim$0.85\,km\,s$^{-1}$) and slightly worse with 0.5\,MHz
($\sim$1.63\,km\,s$^{-1}$) for CH$_3$CN because we had to cover a
broader bandwidth due to the several $K$-components. Flux calibration
was performed with observations of Uranus and is estimated to be
accurate within 20\%. The primary beam of the ATCA at the observing
frequency is $36''$ (FWHM).  The data were reduced with the MIRIAD
package.  Applying different weightings for the line and continuum
data -- mostly uniform weighting for the compact continuum and
CH$_3$CN emission and natural for the more extended HCN emission --
the synthesized beams vary between the different maps. The achieved
spatial resolution and $1\sigma$ rms values are given in Table
\ref{rms}.

\begin{table}[htb]
\caption{Synthesized beams $\theta$ and rms}
\begin{tabular}{lrr}
\hline \hline
Line & $\theta$ & $1\sigma$ rms \\
     & [$''$] & [$\frac{\rm{mJy}}{\rm{beam}}$]\\
\hline
NGC6334I \\
\hline
3.4mm cont.                      & $2.3''\times 1.6''$ & 23 \\
CH$_3$CN$(5_4-4_4)$, [-12,2]km/s$^a$ & $2.2''\times 1.6''$ & 7 \\
CH$_3$CN$(5_4-4_4)$, 1.7km/s$^b$ & $2.2''\times 1.6''$ & 11 \\
HCN(1--0), [-25,-19]km/s$^a$         & $2.6''\times 1.8''$ & 19 \\
HCN(1--0), [5,19]km/s$^a$         & $2.6''\times 1.8''$ & 16 \\
HCN(1--0), 1.7km/s$^b$           & $2.5''\times 1.8''$ & 33 \\
\hline      
NGC6334I(N)\\
\hline
3.4mm cont.                      & $2.6''\times 1.8''$ & 2.6 \\
CH$_3$CN$(5_{k=0,1}-4_{k=0,1})$$^a$, [-12,8]km/s & $2.2''\times 1.6''$ & 4.9 \\
CH$_3$CN$(5_4-4_4)$, 1.7km/s$^b$ & $2.2''\times 1.6''$ & 8.4 \\
HCN(1--0), [-16,-6]km/s$^a$         & $2.5''\times 1.8''$ & 19 \\
HCN(1--0), [4,12]km/s$^a$            & $2.5''\times 1.8''$ & 15 \\
HCN(1--0), 2.0km/s$^b$           & $2.5''\times 1.8''$ & 26 \\
\hline \hline
\end{tabular}
~\\
{\footnotesize $^a$ Integrated velocity regime.\\
$^b$ Channel maps with the given velocity resolution.}
\label{rms}
\end{table}

\section{Results and Discussion}

\begin{figure*}[htb]
\includegraphics[angle=-90,width=17.6cm]{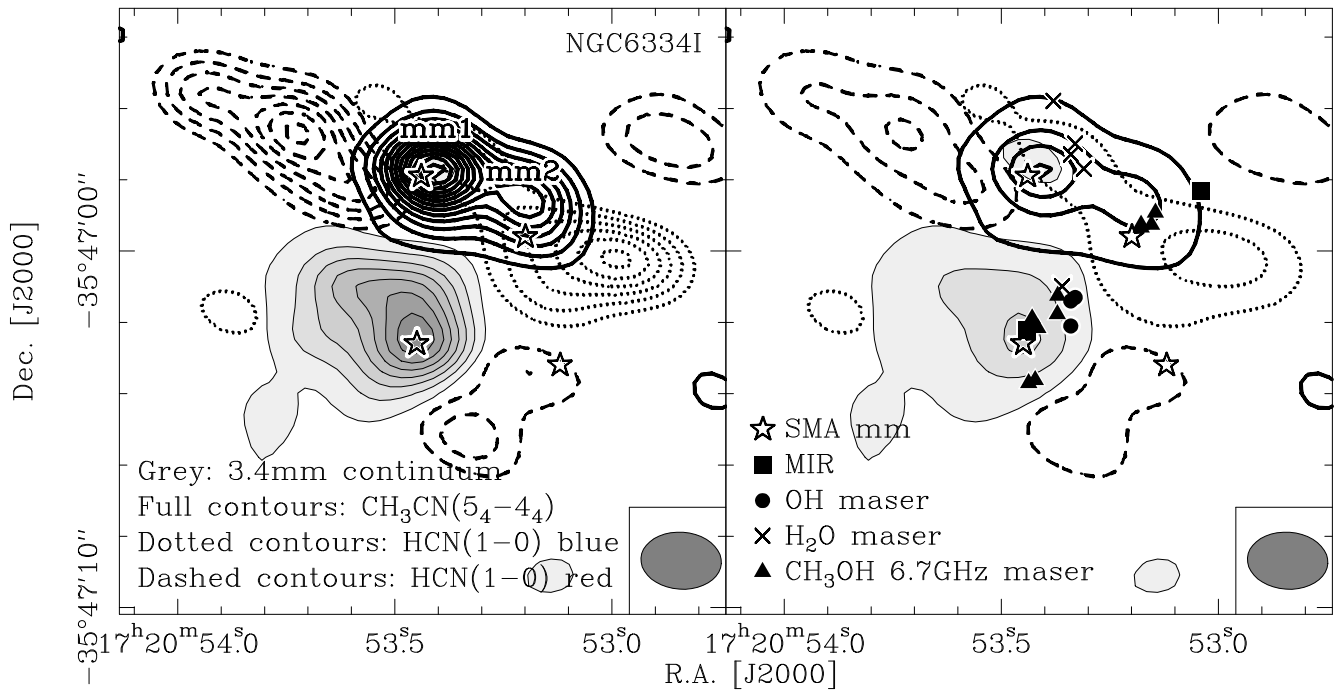}
\caption{3.4\,mm spectral line and continuum emission toward NGC6334I:
  The grey-scale shows the 3.4\,mm continuum and the full contours
  present the integrated CH$_3$CN$(5_4-4_4)$ emission from -12 to
  2\,km\,s$^{-1}$. The dashed and dotted contours show the blue- and
  red-shifted HCN(1--0) emission with integration ranges [-25,-19] and
  [5,19]\,km\,s$^{-1}$, respectively. In the left panel, the 3.4\,mm
  and CH$_3$CN emission is contoured from $3\sigma$ and continue in
  $3\sigma$ steps ($1\sigma(3.4\rm{mm})\sim 23$\,mJy\,beam$^{-1}$ and
  $1\sigma(\rm{CH_3CN})\sim 7$\,mJy\,beam$^{-1}$). The red- and
  blue-shifted HCN contours start at $2\sigma$ and continue in
  $1\sigma$ steps ($1\sigma(\rm{HCN-red})\sim 19$\,mJy\,beam$^{-1}$
  and $1\sigma(\rm{HCN-blue})\sim 16$\,mJy\,beam$^{-1}$). The mm
  sources from \citet{hunter2006} are marked by stars. For clarity,
  the right panel shows the same data but with less contours, and
  additional poitions from various other observations are included
  (identified at the bottom-left). The corresponding references are:
  SMA mm continuum data from \citet{hunter2006}, H$_2$O masers from
  \citet{forster1989}, CH$_3$OH class {\sc ii} masers from
  \citet{walsh1998}, OH masers from \citet{brooks2001}, MIR sources
  from \citet{debuizer2002}). The NH$_3$(6,6)/(8,6)/(11,9) are not
  shown but are spatially associated with mm2
  \citep{beuther2007b,walsh2007}. The synthesized beam is shown at the
  bottom right of each panel.}
\label{ngc6334i}
\end{figure*}

\begin{figure*}[htb]
\includegraphics[angle=-90,width=17.6cm]{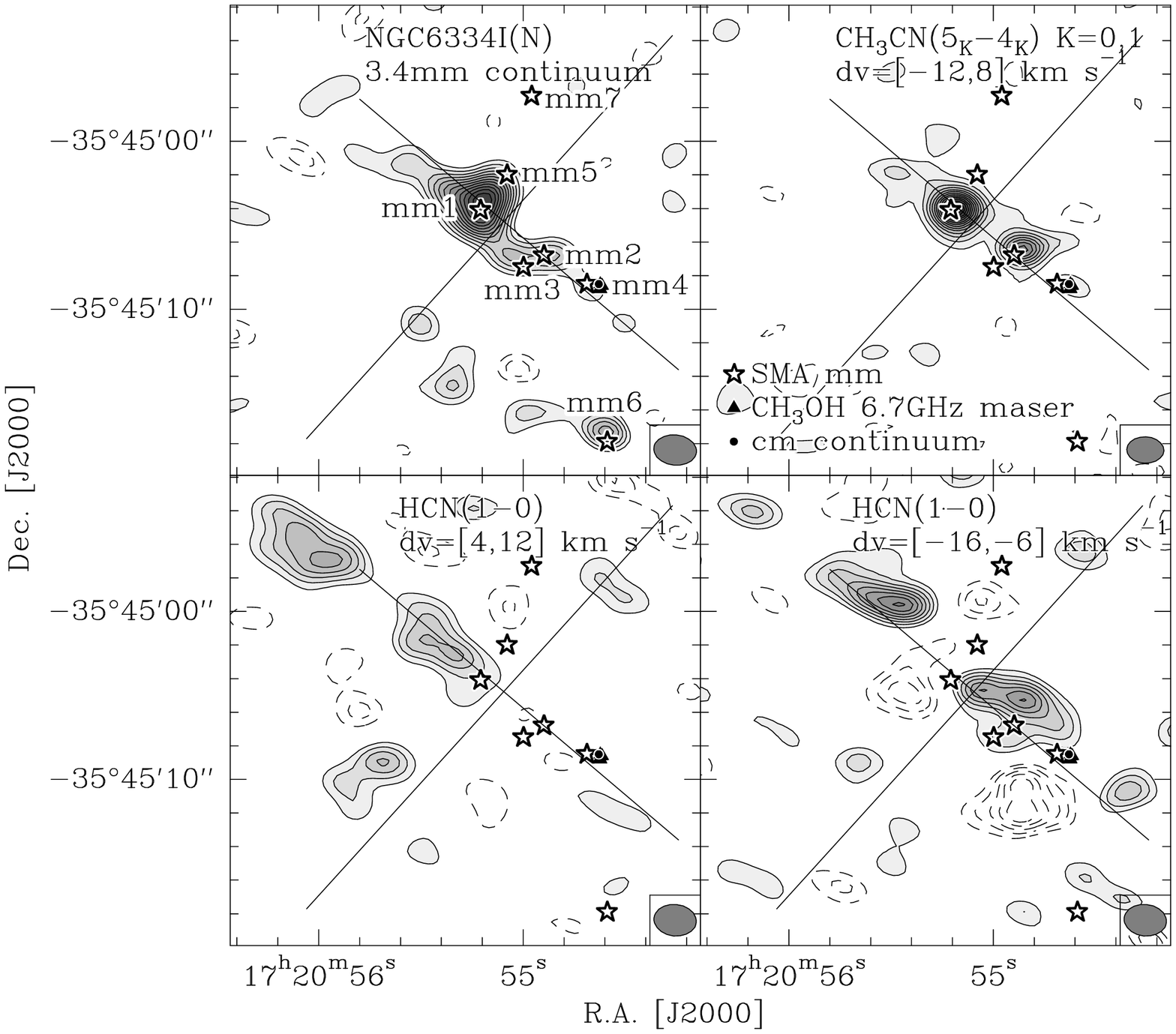}
\caption{3.4\, mm spectral line and continuum emission toward
  NGC6334I(N) The four panels show the continuum (top-left), the
  integrated CH$_3$CN$(5_K-4_K)$ ($K=0,1$, top-right), and the red-
  and blue-shifted HCN(1--0) emission (bottom-left and bottom-right
  respectively). The velocity regimes are given in the figure.  The
  3.4\,mm and CH$_3$CN emission is contoured from $2\sigma$ and
  continue in $2\sigma$ steps ($1\sigma(3.4\rm{mm})\sim
  2.6$\,mJy\,beam$^{-1}$ and $1\sigma(\rm{CH_3CN})\sim
  4.9$\,mJy\,beam$^{-1}$). The red- and blue-shifted HCN wing contours
  start at $2\sigma$ and continue in $1\sigma$ steps
  ($1\sigma(\rm{HCN-red})\sim 19$\,mJy\,beam$^{-1}$ and
  $1\sigma(\rm{HCN-blue})\sim 15$\,mJy\,beam$^{-1}$). Negative
  features caused by insufficient uv-coverage are shown in dashed
  contours with the same levels as the emission. Markers of various
  other observations are identified in the top-right panel (SMA mm
  data from \citet{hunter2006}, the sources are labeled in the
  top-left panel, CH$_3$OH class {\sc ii} maser from
  \citet{walsh1998}, cm emission from \citet{carral2002}). The two
  lines outline the axes of the two molecular outflows identified by
  \citet{megeath1999} Hunter et al.~(in prep.). The synthesized beams
  are shown at the bottom right of each panel.}
\label{ngc6334in}
\end{figure*}

All of the targetted spectral lines with excitation levels above
ground, $E_{\rm{u}}/k$, between 4 and 128\,K (Table \ref{lines}) and
the 3.4\,mm continuum emission were detected and mapped toward both
target regions (Figs.~\ref{ngc6334i} \& \ref{ngc6334in}). While
CH$_3$CN is a typical hot core molecule and only detected toward the
central warm protostars, HCN exhibits significantly more extended
emission and shows a strong association with the various molecular
outflows in the two regions. This may be considered surprising since
the critical density of HCN(1--0) is even larger than that of
CH$_3$CN.  The 3.4\,mm continuum emission is of different origin in
both regions: while it is largely due to the free-free emission from
the UCH{\sc ii} region in NGC6334I, toward the northern region
NGC6334I(N) the continuum emission stems mainly from the dust in the
vicinity of the embedded protostars. In the following we will outline
the characteristics of both regions separately.

\subsection{NGC6334I}

\subsubsection{3.4\,mm continuum emission}

Figure \ref{ngc6334i} presents an overlay outlining the main features
of the spectral line and continuum emission toward NGC6334I. Almost
all of the 3.4\,mm continuum emission arises from the UCH{\sc ii}
region with a comparable morphology to the previous cm continuum
images of that region (e.g., \citealt{depree1995,beuther2005e}). In
contrast to that, it is only barely detectable at a $3.8\sigma$ level
of 87\,mJy\,beam$^{-1}$ toward the strongest 1.4\,mm continuum,
NH$_3$ and CH$_3$CN peak position, mm1 in Fig.~\ref{ngc6334i}
\citep{hunter2006,beuther2007b}, marking the location of the
dominating protostar in the region. Comparing this $3.8\sigma$
detection with the 1.4\,mm data-point from \citet{hunter2006} of
2.09\,Jy\,beam$^{-1}$ (the 1.4\,mm data were re-imaged with the same
beam-size as the 3.4\,mm data), we get a spectral index of $\sim$3.4.
However, since our detection is barely above the $3\sigma$ level, and
mm1 may well harbor a so far undetected hypercompact H{\sc ii} region
that could contribute still significant flux at 3.4\,mm (e.g.,
\citealt{beuther2007c}), we refrain from further analysis of that
feature.

\subsubsection{The dense gas observed in CH$_3$CN$(5_K-4_K)$}
\label{6334i_ch3cn}

\begin{figure}[htb]
\includegraphics[angle=-90,width=8.8cm]{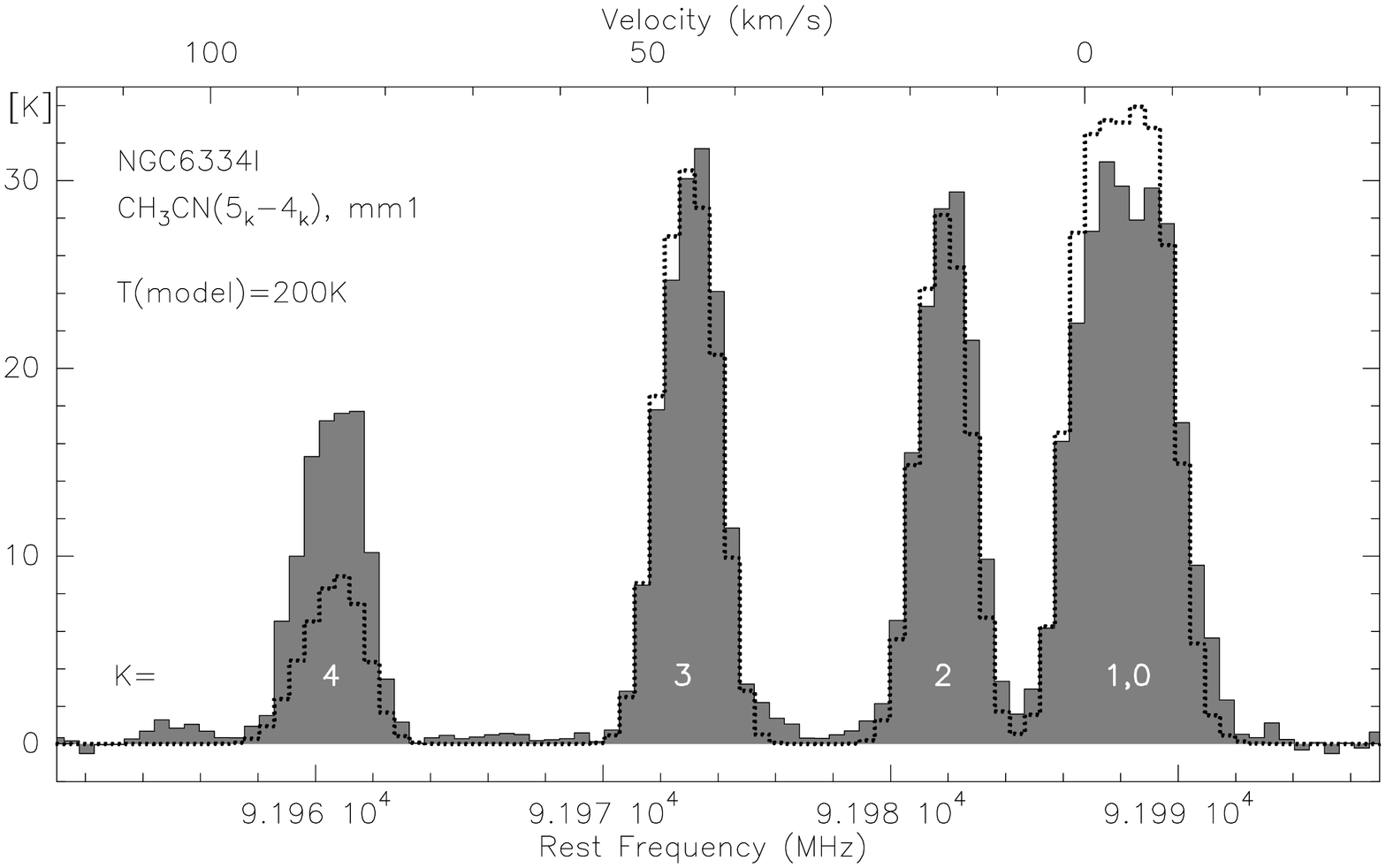}\\
\includegraphics[angle=-90,width=8.8cm]{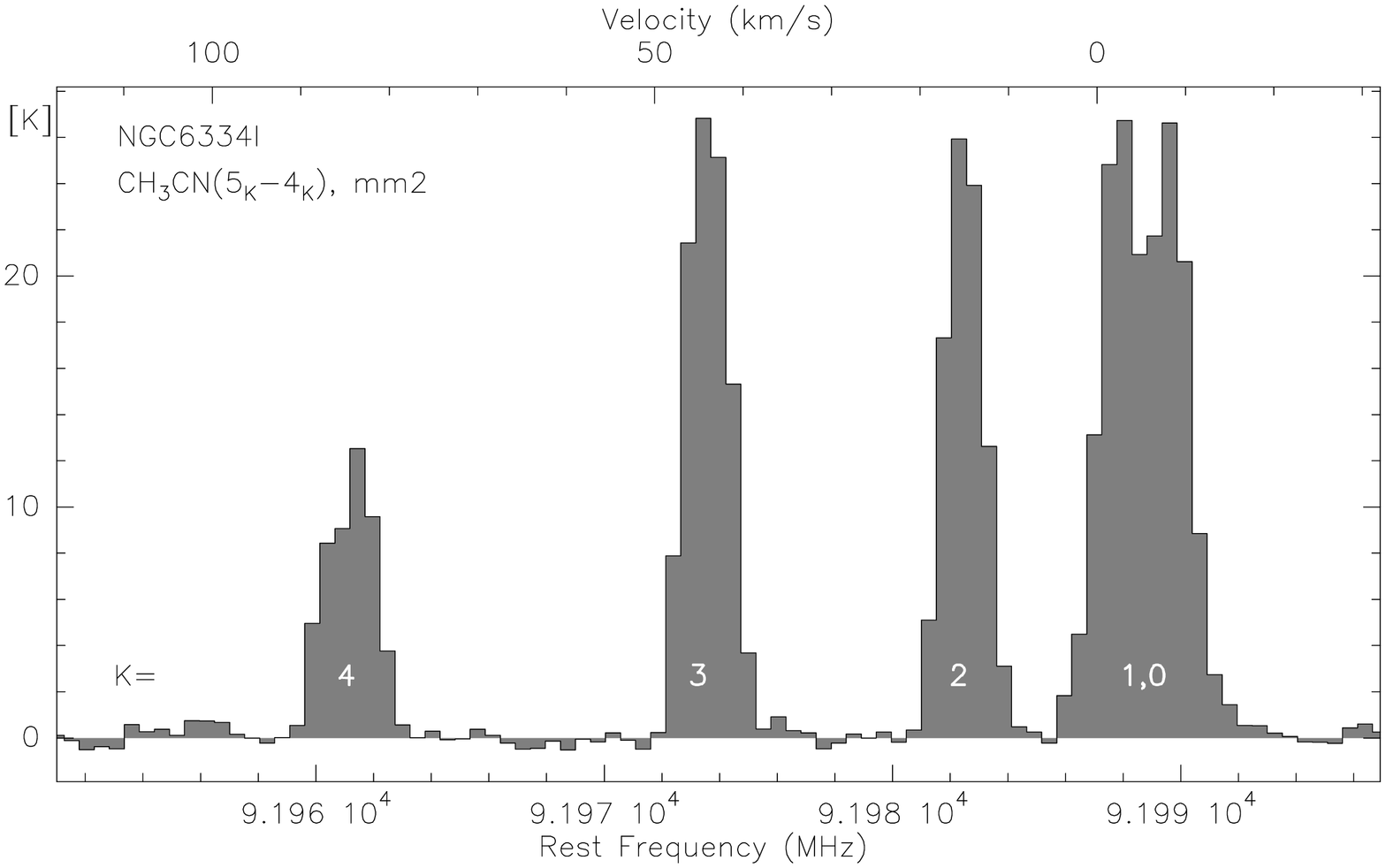}
\caption{CH$_3$CN$(5_K-4_K)$ spectra ($K=0...4$) extracted toward the
  two CH$_3$CN peak positions in NGC6334I shown in Figure
  \ref{ngc6334i}. The dotted line in the top panel shows a model
  spectrum created with XCLASS at a temperature of 200\,K.}
\label{ngc6334i_ch3cn}
\end{figure}

In contrast to the 3.4\,mm continuum emission, the CH$_3$CN emission
distribution shows the typical double-peaked morphology known from the
previous NH$_3$ observations \citep{beuther2005e,beuther2007b}. The
two CH$_3$CN peaks appear to be associated with the two strongest mm
continuum sources (mm1 and mm2) in the region \citep{hunter2006}.
Figure \ref{ngc6334i_ch3cn} presents the full CH$_3$CN$(5_K-4_K)$
spectra extracted toward the two peak positions, and clearly all five
$K$-components up to $K=4$ with $E_{\rm{u}}/k=128$\,K are well
detected.  Table \ref{ch3cn} lists the fitted line parameters of the
spectra.

\begin{table}[htb]
\caption{Fitted CH$_3$CN$(5_K-4_K)$ line parameters}
\begin{tabular}{lrr}
\hline \hline
Line & $S_{\rm{peak}}$ & $\Delta v^b$ \\
     & [K] & [km/s]\\
\hline
NGC6334I, mm1 \\
\hline
CH$_3$CN$(5_0-4_0)^a$ & 24.1 & 7.0$\pm 0.4$ \\
CH$_3$CN$(5_1-4_1)^a$ & 27.8 & 8.3$\pm 0.4$ \\
CH$_3$CN$(5_2-4_2)$   & 29.9 & 7.5$\pm 0.2$ \\
CH$_3$CN$(5_3-4_3)$   & 31.9 & 7.8$\pm 0.03$ \\
CH$_3$CN$(5_4-4_4)$   & 18.7 & 8.5$\pm 0.1$ \\
\hline      
NGC6334I, mm2 \\
\hline
CH$_3$CN$(5_0-4_0)^a$ & 25.6 & 5.1$\pm 0.3$ \\
CH$_3$CN$(5_1-4_1)^a$ & 26.9 & 5.3$\pm 0.3$ \\
CH$_3$CN$(5_2-4_2)$   & 27.3 & 5.3$\pm 0.1$ \\
CH$_3$CN$(5_3-4_3)$   & 28.8 & 5.8$\pm 0.1$ \\
CH$_3$CN$(5_4-4_4)$   & 11.9 & 6.9$\pm 0.2$ \\
\hline      
NGC6334I(N), mm1 \\
\hline
CH$_3$CN$(5_0-4_0)^a$ & 7.4 & 8.3$\pm 0.6$ \\
CH$_3$CN$(5_1-4_1)^a$ & 6.9 & 8.4$\pm 0.8$ \\
CH$_3$CN$(5_2-4_2)$   & 8.7 & 7.6$\pm 0.5$ \\
CH$_3$CN$(5_3-4_3)$   & 8.9 & 7.7$\pm 0.2$ \\
CH$_3$CN$(5_4-4_4)$   & 4.4 & 10.0$\pm 0.4$ \\
\hline      
NGC6334I(N), mm2 \\
\hline
CH$_3$CN$(5_0-4_0)^a$ & 5.6 & 6.6$\pm 1.0$ \\
CH$_3$CN$(5_1-4_1)^a$ & 5.7 & 3.9$\pm 0.5$ \\
CH$_3$CN$(5_2-4_2)$   & 4.6 & 6.5$\pm 0.5$ \\
CH$_3$CN$(5_3-4_3)$   & 4.5 & 6.0$\pm 0.3$ \\
CH$_3$CN$(5_4-4_4)$   & 1.0 & 10.9$\pm 1.6$ \\

\hline \hline
\end{tabular}
~\\
{\footnotesize $^a$ The fits to the $K=0,1$ lines are less accurate because of the strong line blending of both components.}\\
{\footnotesize $^b$ Full Width Half Maximum (FWHM)}\\
\label{ch3cn}
\end{table}

Since the early work by \citet{loren1984} CH$_3$CN has often been used
as a thermometer to estimate rotation temperatures of the dense gas
via Boltzmann plots assuming optically thin emission in Local
Thermodynamic Equilibrium (LTE). We tried this approach here as well,
however, it failed because CH$_3$CN$(5_K-4_K)$ is optically thick.
Similarly, we tried to model the CH$_3$CN spectra in LTE using the
XCLASS superset to the CLASS software developed by Peter Schilke
(priv.~comm., see also \citealt{comito2005}). This software package
uses the line catalogs from JPL and CDMS
\citep{poynter1985,mueller2001}. However, this approach failed as
well. Fig.~\ref{ngc6334i_ch3cn} shows a model spectrum at $T=200$\,K
overlaid on the CH$_3$CN$(5_K-4_K)$ toward the main mm peak mm1.
While the $K=2,3$ components still fit relatively well, the model does
neither reproduce the $K=0,1$ nor the $K=4$ component.  While this is
partly again an opacity effect, it also shows that LTE is not
appropriate for a source like NGC6334I. As outlined below, different
$K$-levels exhibit different line-widths and hence do not trace the
same gas components. Temperature gradients within the sources are
imprinted in the spectra further complicating single-temperature fits.
More sophisticated modeling of the CH$_3$CN emission is warranted.
Although a few radiative transfer calculations of CH$_3$CN data exist
(e.g., \citealt{olmi1993}), the sparsely available collisional
transition rates make such calculations a difficult task (the only
partly available data are a small compilation by
\citealt{pei1995a,pei1995b}). Furthermore, in such hot and dense
regions radiative excitation starts to matter which is hard to account
for in any modeling approach.

\begin{figure}[htb]
\includegraphics[angle=-90,width=8.8cm]{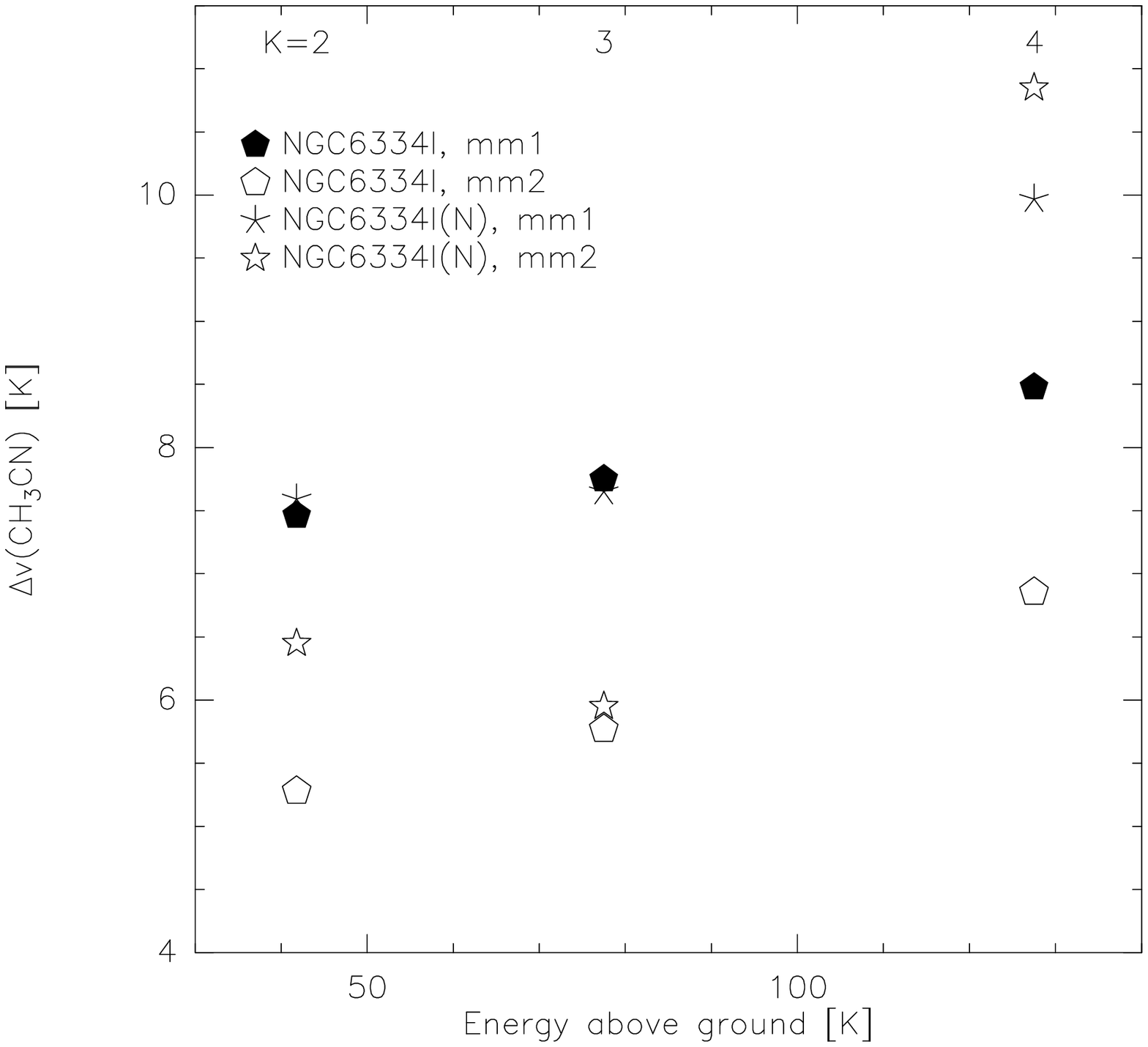}
\caption{Line-widths $\Delta v(\rm{CH_3CN})$ from the various
  $K$-components plotted against the upper energy level of each line.
  We omitted the $K=0,1$ lines because of the line blending.  The
  different symbols correspond to 4 different peak positions in
  NGC6334I and I(N) as labeled in the plot. The corresponding errors
  are given in Table \ref{ch3cn}.}
\label{dv}
\end{figure}

\begin{figure}[htb]
\includegraphics[angle=-90,width=8.8cm]{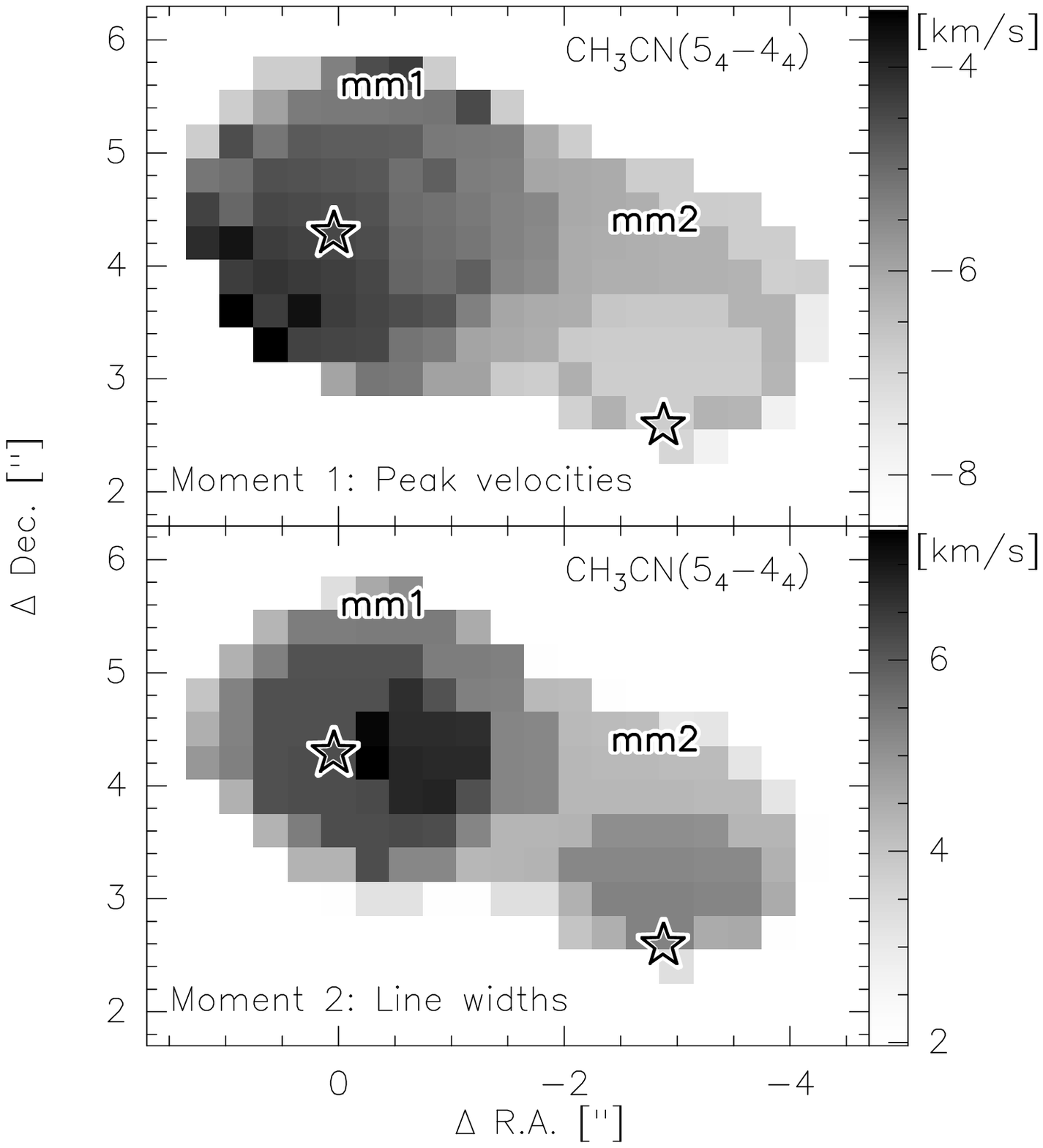}
\caption{Moment maps of CH$_3$CN$(5_4-4_4)$ toward NGC6334I. The top
  panel shows the 1st moment (peak velocities) and the bottom panel
  the 2nd moment (line widths). The stars mark the positions of the
  two main mm continuum sources from \citet{hunter2006}. In addition
  to the velocity gradient from between mm1 and mm2, the 1st moment
  map exhibits a 2nd velocity gradient around mm1 in north-west
  south-east direction (approximately perpendicular to the large-scale
  outflow).}
\label{mom_ch3cn_6334i}
\end{figure}

While the Full Width Half Maximum (FWHM) line-widths $\Delta v$ for
the two lowest energy lines of the $K=0,1$ components are less certain
because of the line-blending between both components, one identifies a
trend of increasing line-width, $\Delta v$, with increasing $K$
quantum number for $K\geq 2$ (Table \ref{ch3cn}). Figure \ref{dv}
shows the $\Delta v$ of CH$_3$CN$(5_K-4_K)$ for $K\geq 2$ toward the
two CH$_3$CN peaks in NGC6334I \& I(N) plotted versus the level energy
above ground, $E_{\rm{u}}/k$. For all four positions the trend of
increasing $\Delta v$ versus $E_{\rm{u}}/k$ is discernable. This trend
indicates more internal motions from the warmer gas components (traced
by the higher $E_{\rm{u}}/k$ lines) which perhaps originate from the
inner warm regions close to the protostars.  Similar to that, the
bottom-panel of Figure \ref{mom_ch3cn_6334i} shows the 2nd moment map
of the CH$_3$CN$(5_4-4_4)$ line, i.e., its line-width distribution.
Again we see the line-width increase toward the center close to the
main mm continuum peaks. Different processes may cause such line
broadening, e.g., accretion disk rotation, infall or outflow motions.
In particular, accretion disks are interesting candidates to explain
such observational features.

\begin{figure}[htb]
\includegraphics[angle=-90,width=8.8cm]{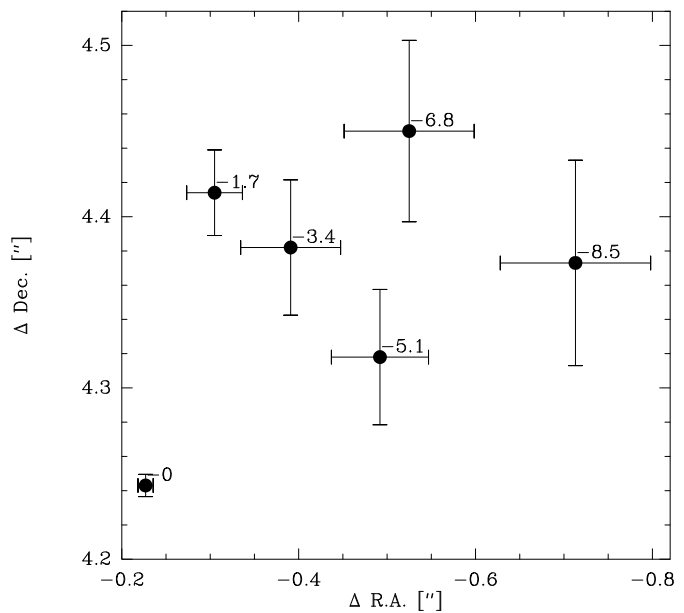}
\caption{Positional offsets around mm1 of the different CH$_3$CN
  $(5_4-4_4)$ velocity channels derived via Gaussian fits of the peak
  emission in each separate channel. The error-bars are the nominal
  $1\sigma$ errors from the fits, and the numbers label the central
  velocity for each position.}
\label{ch3cn_offsets}
\end{figure}

To investigate rotation from a potentially embedded massive accretion
disk, the top-panel of Figure \ref{mom_ch3cn_6334i} presents the 1st
moment map of the CH$_3$CN$(5_4-4_4)$ line, i.e., its peak velocity
distribution. Although this structure is only barely resolved by the
synthesized beam of $2.5''\times 1.8''$ we tentatively identify a
velocity gradient across mm1 with an approximate position angle (PA)
of $113\pm 23$ degrees from north. The same velocity structure is
descernable in the lower $K=3,2$ CH$_3$CN transitions (we refrained
from analyzing the $K=0,1$ lines because of their line-blending).  To
investigate this potential velocity gradient in more detail, we fitted
the peak positions of each independent spectral channel
(Fig.~\ref{ch3cn_offsets}) which should allow us to increase the
resolving power to approximately 0.5\,HPBW/(S/N), where HPBW equals
the synthesized beam and S/N the signal-to-noise ratio
\citep{reid1988}. Similar to the moment map, the case is not
clear-cut, but nevertheless, the data are indicative of a velocity
gradient with a PA of $\sim 134^{+20}_{-37}$ degrees from north. In
comparison to these position angles, the PA of the previously
identified molecular outflow is $\sim 46$ degrees from north
\citep{bachiller1990}, which is approximately perpendicular to that
found from our measurements, especially those from the
highest-spatial-resolution position fitting
(Fig.~\ref{ch3cn_offsets}).

\begin{figure}[htb]
\includegraphics[angle=-90,width=8.8cm]{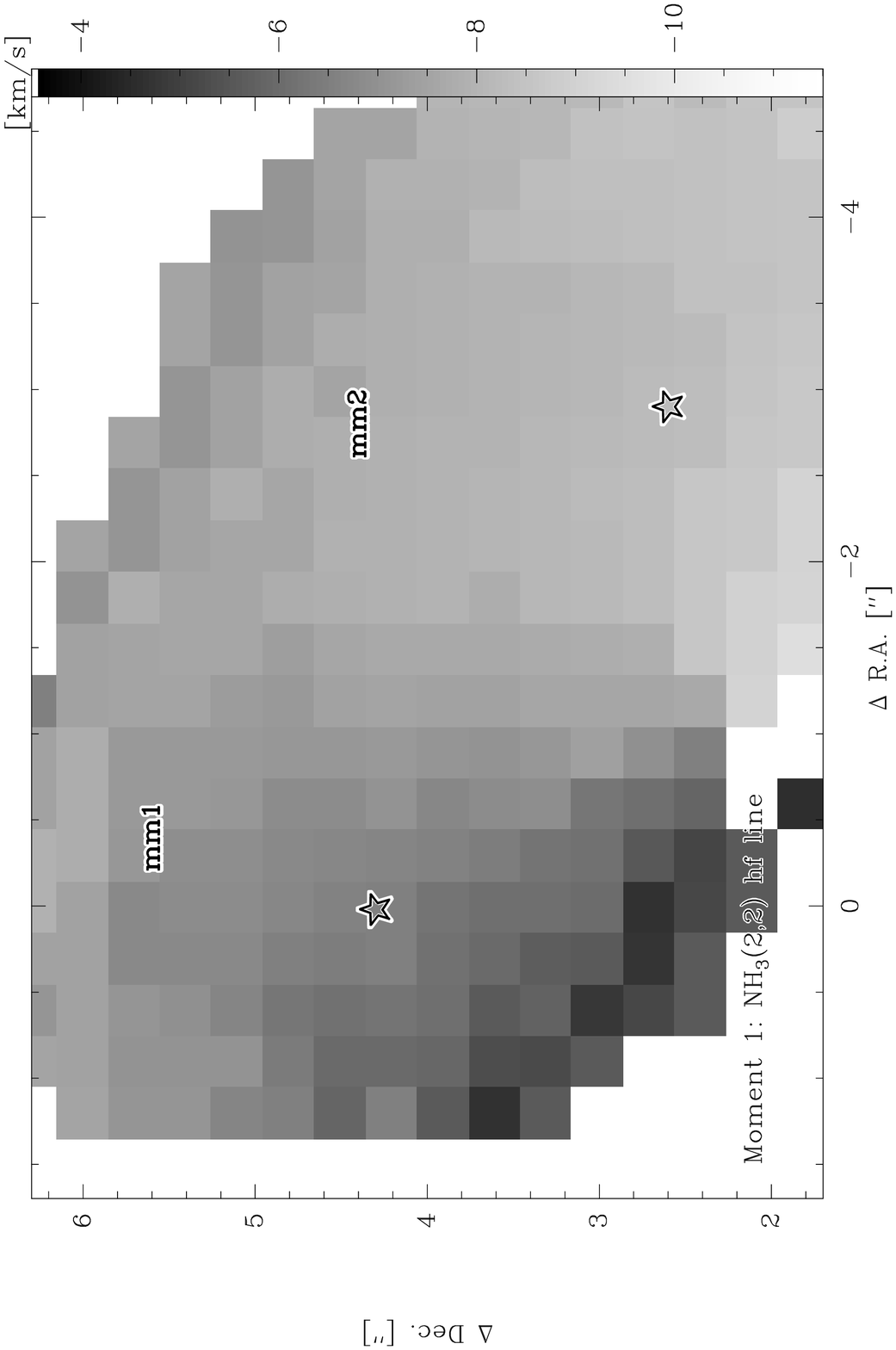}
\includegraphics[angle=-90,width=8.8cm]{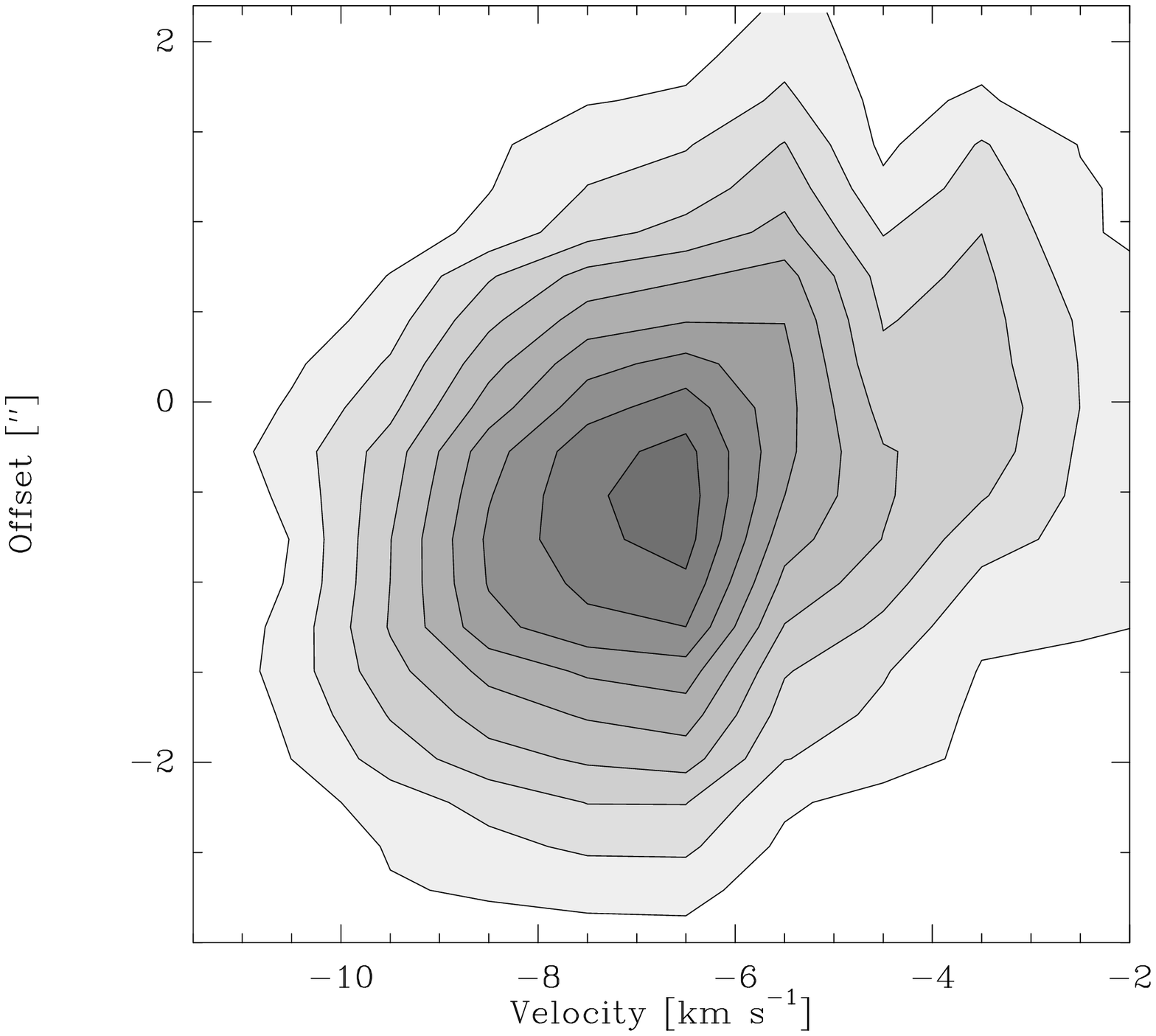}
\caption{First moment map (top) and position velocity diagram (bottom)
  of the the most blue-shifted satellite hyperfine line of NH$_3$(2,2)
  toward NGC6334I. The data are re-examined from \citet{beuther2005e}.
  The position velocity digram is centered on mm1 with a position
  angle of 136 degrees from north. In addition to the velocity
  gradients between mm1 and mm2, the data exhibit a 2nd velocity
  gradient around mm1 in north-west south-east direction
  (approximately perpendicular to the large-scale outflow).  The
  synthesized beam is $2.8''\times2.2''$.}
\label{mom1_nh3_6334i}
\end{figure}

Based on these findings, we went back to the previous NH$_3$(1,1) to
(6,6) observations \citep{beuther2005e,beuther2007b} searching for
similar signatures in these data. While the main hyperfine lines as
well as the satellite lines of NH$_3$(1,1) are dominated by the
large-scale velocity gradient over the two main cores (e.g., Fig.~13
in \citealt{beuther2005e}), the satellite lines of the (J,K) lines
with J,K$\geq$3 are overlapping and difficult to image. However, the
satellite hyperfine lines of the NH$_3$(2,2) transition exhibit, in
addition to the velocity gradient over the two cores, a second
velocity gradient across mm1, again approximately perpendicular to the
large-scale outflow (Fig.~\ref{mom1_nh3_6334i}).  The confirmation of
this velocity gradient, first identified in the CH$_3$CN$(5_4-4_4)$
line, now in the lower opacity satellite hyperfine NH$_3$(2,2) line
supports its general credibility.

Since we do not resolve well the substructure of that velocity
gradient to better investigate the kinematics (e.g., is there any
Keplerian motion present?) the data do not allow the claim that a
massive accretion disk was detected. For example, an unresolved
double-source with different line-of-sight velocities could produce a
similar signature (e.g., \citealt{brogan2007}). Nevertheless, these
observations are suggestive of a rotating structure perpendicular to
the molecular outflow within a projected diameter of $\sim 0.33''$
derived from Fig.~\ref{ch3cn_offsets}. At the given distance of
1.7\,kpc, this corresponds to a radius of the rotating structure of
$\sim$280\,AU.  This scale fits well to the sizes of the accretion
disks simulated recently via 3-dimensional radiative-transfer
hydrodynamic calculations by \citet{krumholz2006b}.

Adopting the proposed disk scenario, we can estimate the approximate
rotationally supported binding mass $M_{\rm{rot}}$ assuming
equilibrium between the centrifugal and gravitational forces at the
outer radius of the disk. Then we get

\begin{eqnarray}
M_{\rm{rot}} & = & \frac{\delta v^2r}{G} \label{eq1} \\
\Rightarrow M_{\rm{rot}}[\rm{M_{\odot}}] & = &  1.13\,10^{-3} \times \delta v^2[\rm{km/s}]\times r[\rm{AU}] \label{eq2}
\end{eqnarray}

\noindent $r$ is the disk radius, and $\delta v$ the Half Width Zero
Intensity (HWZI) of the spectral line, approximately 5.1\,km\,s$^{-1}$
(half the velocity range shown in Fig.~\ref{ch3cn_offsets}).  Equations
\ref{eq1} \& \ref{eq2} have to be divided by sin$^2(i)$ where $i$ is
the unknown inclination angle between the disk plane and the plane of
the sky ($i=90^{\circ}$ for an edge-on system). With the given values
we can estimate $M_{\rm{rot}}$ to $\sim 8$/(sin$^2(i)$)\,M$_{\odot}$.
This is of the same order as the mass derived from the mm continuum
emission \citep{hunter2006} which is assumed to stem largely from the
disk/envelope system.

How do these masses correspond to the mass and luminosity of the
central embedded source? While the bolometric luminosity of NGC6334I
is estimated to be $\leq 2.6\times 10^5$\,L$_{\odot}$
\citep{sandell2000}, approximately $0.32\times 10^{5}$\,L$_{\odot}$
are attributed to the UCH{\sc ii} region (based on the Lyman continuum
flux presented in \citealt{depree1995}).  One should keep in mind that
this value may be a lower limit since dust could absorb a significant
fraction of the uv-photons \citep{kurtz1994}. The remaining $\leq
2.3\times 10^5$\,L$_{\odot}$ has to be due to the various sources
associated with the hot molecular core.  Since the associated
mid-infrared source to the west (Fig.~\ref{ngc6334i}) is of relatively
low luminosity (only 67\,L$_{\odot}$, \citealt{debuizer2002}), it is
likely that most of the luminosity stems from the two main continuum
and spectral line peaks. In the extreme case, splitting the luminosity
simply by two, it still implies that about $\leq 1.15\times
10^5$\,L$_{\odot}$ emanate from the mm1 region.  Assuming a ZAMS star,
this corresponds to an embedded star of $\leq$30\,M$_{\odot}$. In the
above adopted accretion disk scenario, this would imply an inclination
angle $i$ between the disk plane and the plane of the sky of
approximately 30 degrees. However, with the large uncertainties in the
rotational mass and the central object mass estimate, such an inclination
angle estimate should not be taken at face value, but only gives a
rough idea that the system is neither edge- nor face-on.  The
comparable mass derived from the mm continuum emission
\citep{hunter2006} indicates that a significant fraction of the
total system mass stems from the proposed accretion disk and envelope.
This implies that the rotating structure is unlikely in Keplerian
motion but that it may be a self-gravitating structure and potential
site of ongoing sub-fragmentation (see also comparable analytic
calculations and hydro-simulations by, e.g., \citealt{kratter2006} and
\citet{krumholz2006b}).

However, on a cautionary note, one has to keep in mind that the data
are not conclusive as to whether we really see disk signatures or
whether the observed gradient may be caused by other motions, e.g., an
unresolved double-source.

\subsubsection{The molecular outflow observed in HCN(1--0)}

\begin{figure}[htb]
\includegraphics[angle=-90,width=8.8cm]{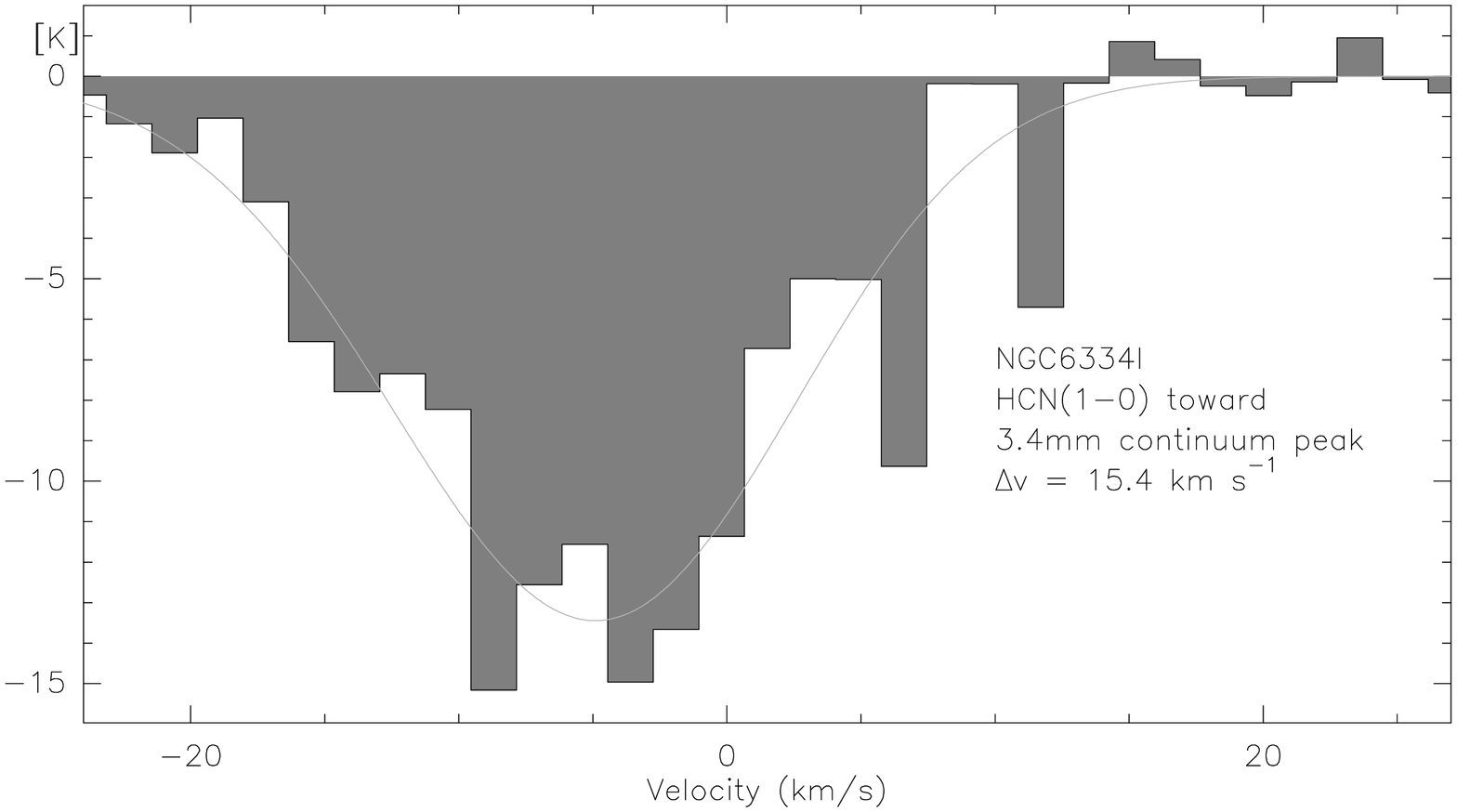}\\
\includegraphics[angle=-90,width=8.8cm]{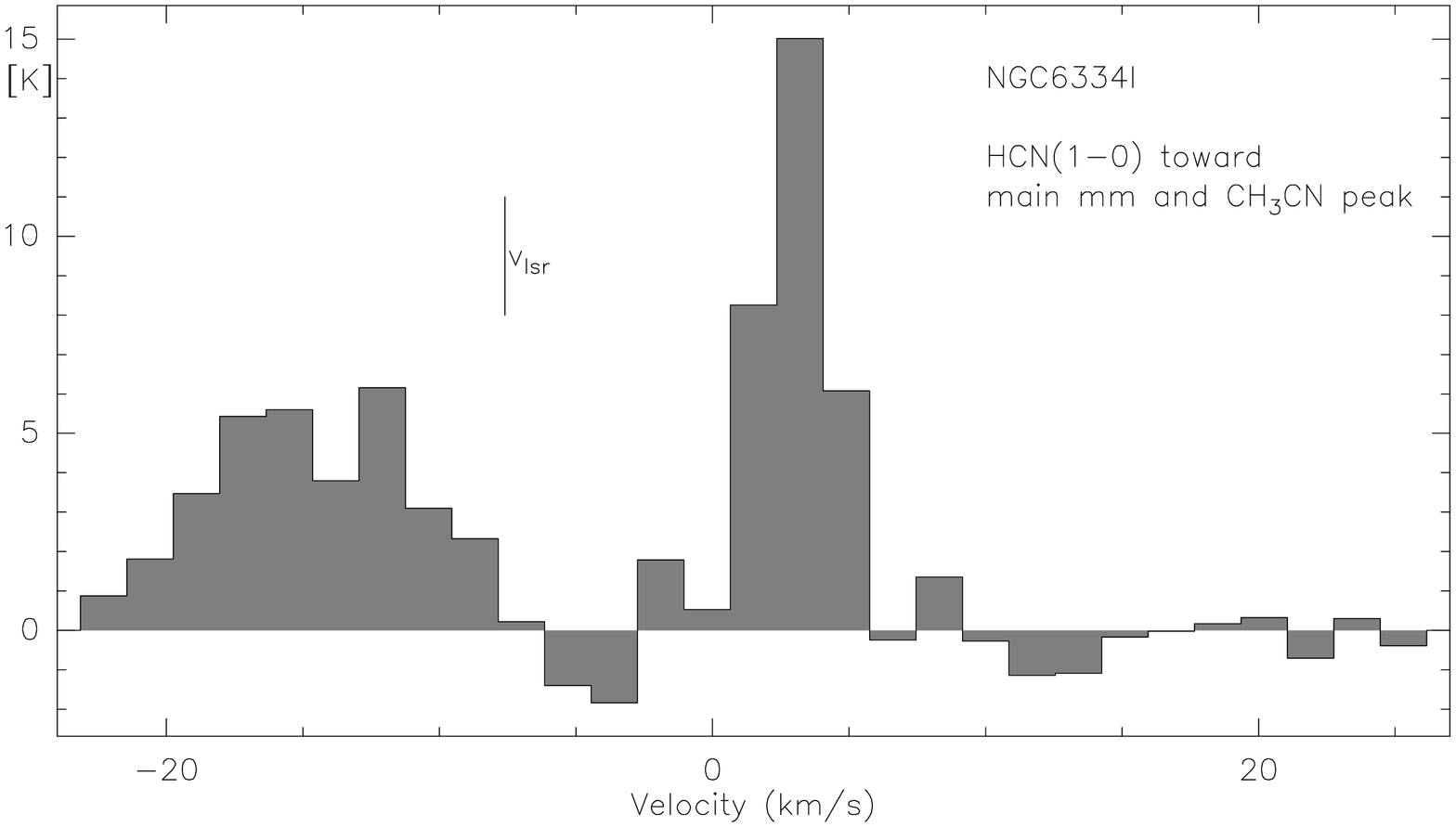}
\caption{HCN(1--0) spectra in NGC6334I. The top spectrum shows the
  strong absorption toward the 3.4\,mm continuum peak that coincides
  with the UCH{\sc ii} region previously observed at cm wavelengths.
  The bottom spectrum is extracted toward the main CH$_3$CN peak that
  coincides with the main mm continuum peak SMA1 by
  \citet{hunter2006}.}
\label{ngc6334i_hcn}
\end{figure}

\begin{figure}[htb]
\includegraphics[angle=-90,width=8.8cm]{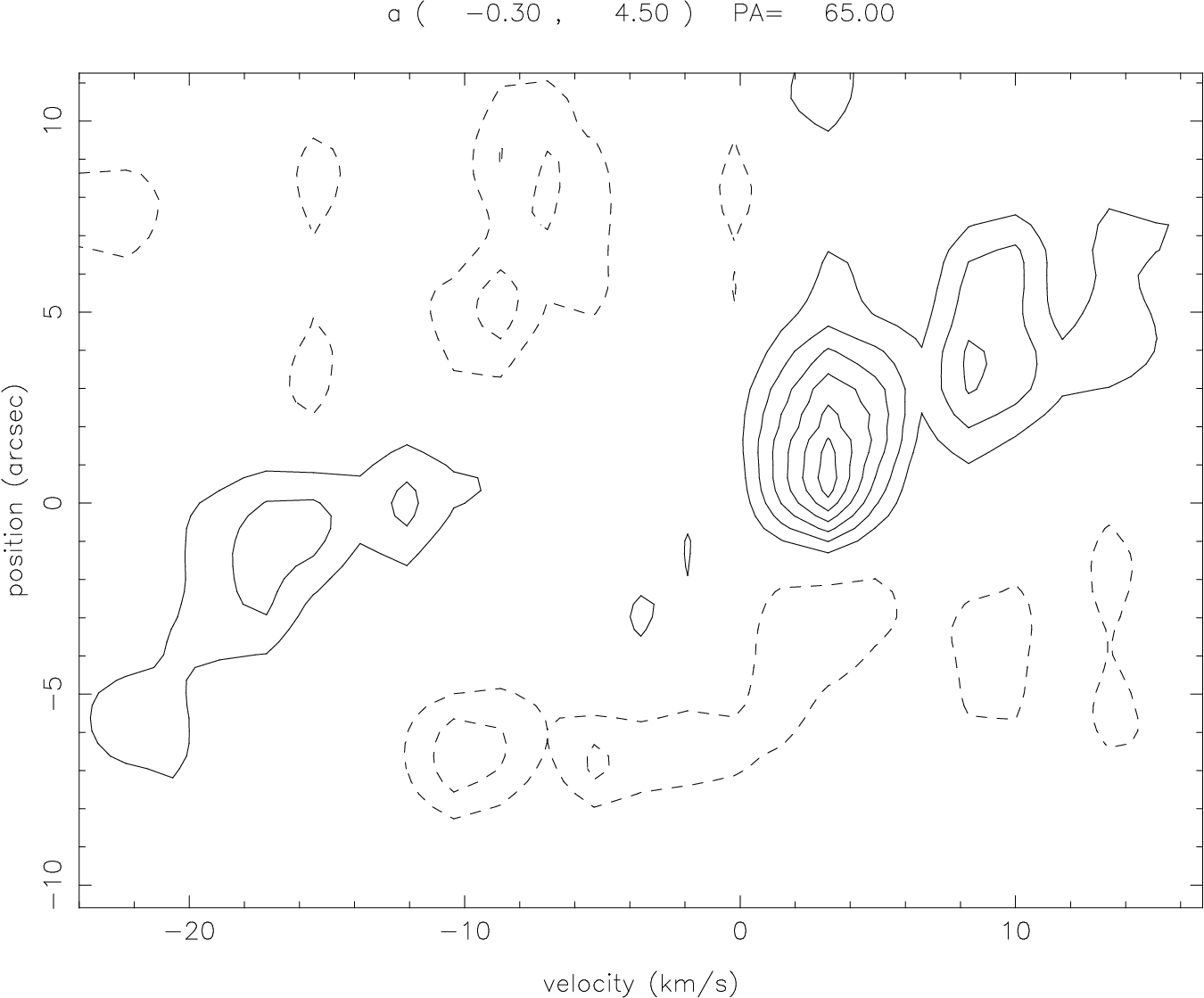}
\caption{Position-velocity diagram of HCN(1--0) in NGC6334I along the
  main outflow axis. The cut is centered on the main CH$_3$CN peak at
  positional offset $-0.3''/4.5''$ with a position angle of
  65$^{\circ}$ from north-to-east. Emission at the velocity of rest
  around $-7.6$\,km\,s$^{-1}$ is largely filtered out.}
\label{pv_hcn_6334i}
\end{figure}

The blue- and red-shifted HCN(1--0) emission in Figure \ref{ngc6334i}
shows high-velocity emission associated with the molecular outflow in
north-east south-western direction \citep{bachiller1990,leurini2006}.
The PA of the emission is not exactly the 45 degrees derived
previously from the single-dish CO observations but it is closer to 65
degrees. However, such a discrepancy is not necessarily a
surprise if one considers the missing flux problem we encounter in the
HCN observations. Figure \ref{ngc6334i_hcn} (bottom panel) presents
the HCN spectrum extracted toward the mm1 peak position, and while we
see well the blue- and red-shifted emission, nearly all the flux
around the $v_{\rm{lsr}}$ of $\sim -7.6$\,km\,s$^{-1}$ is filtered
out. This is also the reason why we do not see HCN emission from the
hot core itself, which is prominent in CH$_3$CN.  Therefore, we just
see some selectively chosen part of the outflow in HCN that could for
example be associated with the limb-brightened cavity walls of the
outflow which would explain the different apparent PA of the image.
Figure \ref{pv_hcn_6334i} shows a position-velocity diagram centered
at the main mm emission and CH$_3$CN peak mm1 along the apparent axis
of the HCN emission. Again we find no emission around the systemic
$v_{\rm{lsr}}$ but going to higher velocities, the blue- and
red-shifted gas exhibits increasing velocity with increasing distance
from the center, closely resembling the typical Hubble-law of
molecular outflows (e.g., \citealt{lee2001}). Such Hubble-law like
behavior can be explained on the smallest jet-scales close to the
protostar by the decreasing gravitational potential of the central
star, however, on the larger scales we observe here this effect gets
negligible and the Hubble-law of molecular outflows is explained by a
density gradient decreasing with distance combined with the continuous
(or episodic) driving of the jet that constantly induces energy in the
outflow (e.g., \citealt{shu1991,smith1997,downes1999}).

\begin{figure*}[htb]
\includegraphics[angle=-90,width=17.6cm]{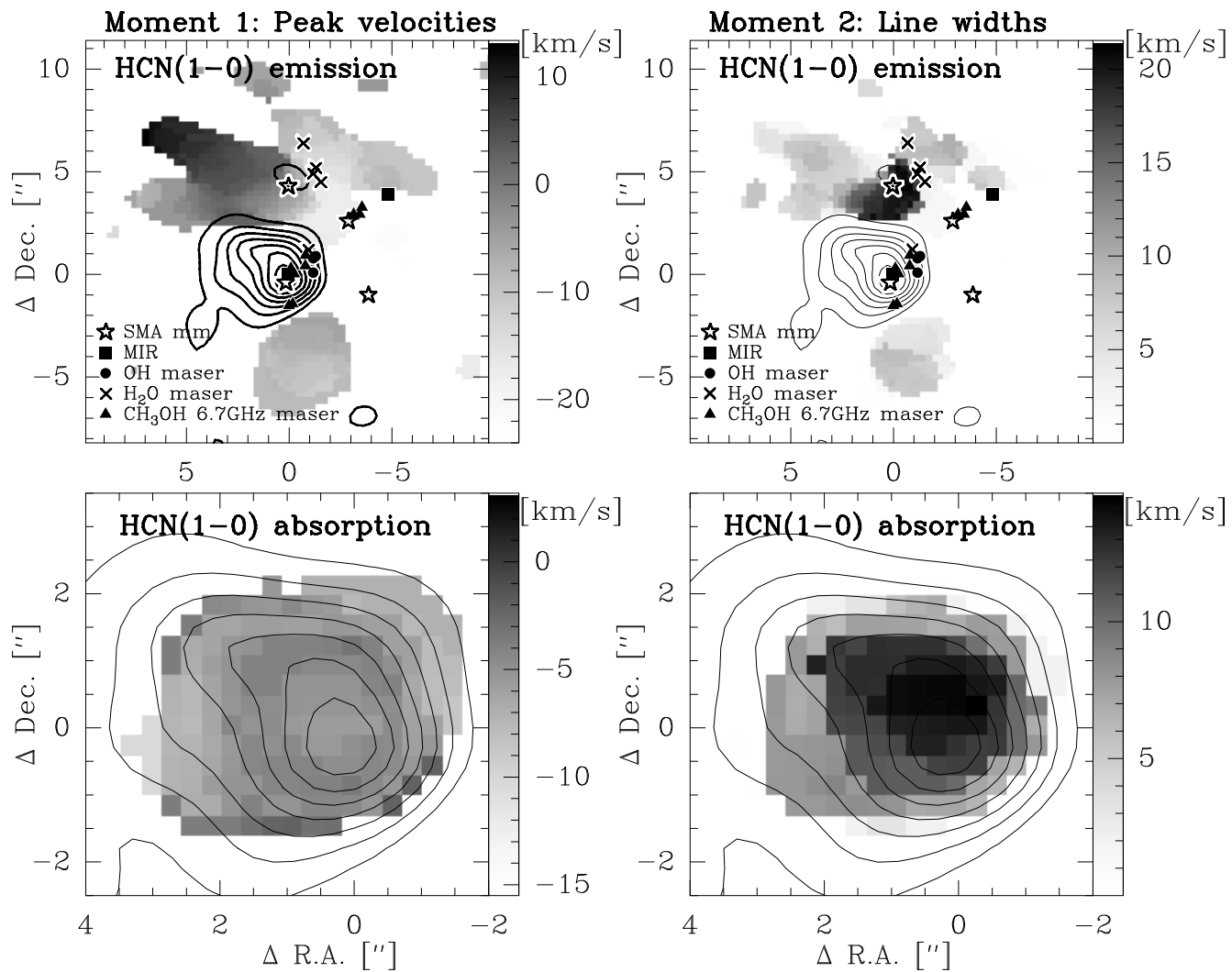}
\caption{Moment maps of HCN(1--0) toward NGC6334I. The left and right
  columns present the first and second moments (peak velocities and
  line widths), respectively. The top-row shows the moments for the
  emission of the outflow and ambient gas, whereas the bottom-row
  presents the corresponding parameters for the absorption features
  toward the UCH{\sc ii} region. The contours present the 3.4\,mm
  continuum emission in $3\sigma$ contour levels ($3\sigma\sim
  69$\,mJy\,beam$^{-1}$). The markers are the same as in
  Fig.~\ref{ngc6334i}.}
\label{mom_hcn_6334i}
\end{figure*}

Figure \ref{mom_hcn_6334i} (top panels) presents the 1st and 2nd
moment map (intensity-weighted peak velocity and line-width
distributions) of the HCN(1--0) emission.  It is interesting to note
that the line-width distribution clearly peaks toward the strongest mm
continuum and molecular line source mm1 and that the large-scale HCN
velocity gradient is approximately centered toward that position as
well. This is indicative of a scenario which puts the driving source
of the molecular outflow observed in HCN at this position.  However,
previous NH$_3$(6,6) maser observations showed the maser peak position
being associated with mm2 with additional features along the axis of
the CO molecular outflow \citep{beuther2007b}.  These features were
interpreted as indicative of this outflow being driven by a source
associated with mm2. While the case is not clear-cut with two
independent and different outflow driver indications, it is also
possible that we are witnessing two different outflows observed in HCN
and CO that may emanate from mm1 and mm2, respectively. In this
scenario, mm1 could be the driver of a potentially denser and younger
outflow that is better detected in HCN, whereas mm2 may be the driver
of the larger, possibly older outflow observed in CO.  Future
observations at higher angular resolution and/or with different
outflow tracers are required to assess the validity of this scenario.

\subsubsection{Absorption toward the UCH{\sc ii} region}

Another interesting feature of the HCN data is that we see strong
absorption in the direction of the UCH{\sc ii} region, similar to the
previously observed absorption in CH$_3$OH and NH$_3$
\citep{beuther2005e}. Figure \ref{ngc6334i_hcn} (top panel) shows the
HCN(1--0) spectrum extracted toward the 3.4\,mm continuum peak
position. Fitting the HCN(1--0) line-width, we take into account its
hyperfine structure consisting of three lines ($F=0-1, 2-1, 1-1$) with
relative intensities of 1:5:3 in the optically thin limit and velocity
shifts of -7.1, 0 and 4.9\,km\,s$^{-1}$, respectively
\citep{poynter1985}. The line-width $\Delta v$ of that absorption
feature is 15.4\,km\,s$^{-1}$, much broader than the previously
observed line-widths between 1.7 and 2.1\,km\,s$^{-1}$ for CH$_3$OH
and between 1.3 and 1.9\,km\,s$^{-1}$ for NH$_3$(1,1) and NH$_3$(2,2)
\citep{beuther2005e}. Table \ref{lines_old} lists the observed
line-widths as well as the upper level excitation temperatures
$E_{\rm{u}}/k$ and the critical densities $n_{\rm{crit}}$ for all
observed absorption lines. The other extreme of the line-widths
distribution is the line-width of the ionized gas observed in the
H76$\alpha$ line $\Delta v=32.0$\,km\,s$^{-1}$ \citep{depree1995}.

Inspecting the values in Table \ref{lines_old} one can discern two
trends: the first is the distinction between the broad line-width and
high critical density line HCN versus the small line-width and low
critical density lines from NH$_3$ and CH$_3$OH.  In addition, within
the low critical density molecules, one can tentatively identify a
correlation between increasing $E_{\rm{u}}/k$ and increasing
line-width.  Although judging from the errors this correlation is less
clear, it remains suggestive.

\begin{table}[htb]
\caption{Line parameters for all absorption lines}
\begin{tabular}{lrrr}
\hline \hline
Line & $E_{\rm{u}}/k$ & $n_{\rm{crit}}$ & $\Delta v$\\
     & [K]          & [cm$^{-3}$]       & [km/s] \\
\hline
NH$_3$(1,1)                 & 24 & 2.1e3 & $1.4\pm 0.4$ \\
NH$_3$(2,2)                 & 65 & 2.2e3 & $2.0\pm 0.3$ \\
CH$_3$OH$(3_{2,1}-3_{1,2})$ & 36 & 1.0e4 & $1.8\pm 0.1$ \\
CH$_3$OH$(4_{2,2}-4_{1,3})$ & 45 & 1.2e4 & $2.1\pm 0.1$ \\
CH$_3$OH$(2_{2,0}-2_{1,1})$ & 29 & 0.8e4 & $1.7\pm 0.1$ \\
HCN(1--0)                   &  4 & 2.6e6 & $15.4\pm 0.3$ \\
\hline \hline
\end{tabular}
~\\
{\footnotesize The NH$_3$ and CH$_3$OH data are from \citet{beuther2005e}. 
  The listed parameters are the upper level excitation energies 
  $E_{\rm{u}}/k$, the critical densities $n_{\rm{crit}}$ calculated at 
  60\,K and the observed line-widths of the absorption lines toward the 
  UCH{\sc ii} region peak position.}
\label{lines_old}
\end{table}

There are different possibilities to explain such trends: The picture
of an expanding UCH{\sc ii} region\footnote{In contrast to the
  previous CH$_3$OH and NH$_3$ absorption lines that showed additional
  blue-shifted emission indicative of expanding gas around the UCH{\sc
    ii} region \citep{beuther2005e}, the HCN data do not exhibit any
  clear expansion or infall signature.} \citep{beuther2005e} in its
surrounding envelope implies that closer to the UCH{\sc ii} region
surface the molecular densities and temperatures are higher than
further outside. Therefore, spectral lines tracing higher densities
around an expanding UCH{\sc ii} region are expected to exhibit broader
line-widths because their emitting gas is more directly impacted than
the lower-density medium further out. A similar explanation could hold
for the different excitation temperature regimes as well. A point of
caution is that the opacity of HCN is probably one to two orders of
magnitude larger than that of, e.g., NH$_3$. While we do not detect
any NH$_3$ satellite hyperfine structure lines in absorption implying
low NH$_3$ opacities, we cannot infer that exactly for HCN. If the
HCN(1--0) optical depth were that high that it could not trace the
dense gas regions close to the expanding UCH{\sc ii} region, then the
above picture could hardly hold. Another explanation is based on the
enhancement of HCN in the molecular outflow. Although the outflow does
not emanate from the UCH{\sc ii} region but from the neighboring mm
continuum source(s), \citet{depree1995} found a velocity gradient in
the ionized gas similar to that of the molecular outflow. They suggest
that the molecular outflow(s) in the region may well disturb the
velocity field of the UCH{\sc ii} region and produce the velocity
gradient this way. Since HCN is known to be abundant in molecular
outflows (see also IRAS\,18566+0408, \citealt{zhang2007}, and
IRAS\,20126+4104, Liu et al.~in prep.) a similar effect could explain
the broad HCN(1--0) line-widths toward the UCH{\sc ii} region.
However, the 1st and 2nd moment maps (peak velocities and line-widths)
of the HCN(1--0) absorption presented in Fig.~\ref{mom_hcn_6334i}
(bottom panels) show no velocity gradient but a peak of the line-width
distribution close to the center of the UCH{\sc ii} region. This is
counter-intuitive for the outflow picture and suggests that the
line-width differences are caused by the expanding UCH{\sc ii} region,
possibly in a fashion comparable to that proposed above. Nevertheless,
without knowing the optical depth of the HCN(1--0) line we cannot
distinguish between the two scenarios.

\subsection{NGC6334I(N)}

\subsubsection{Millimeter continuum emission}

Figure \ref{ngc6334in} (top-left panel) presents the 3.4\,mm continuum
emission toward NGC6334I(N). In contrast to NGC6334I, where we only
see the UCH{\sc ii} region in the 3.4\,mm continuum, toward
NGC6334I(N) we detect four out of seven previously
identified protostellar 1.4\,mm dust continuum condensations
\citep{hunter2006} above a $5\sigma$ level of 13\,mJy\,beam$^{-1}$.
The peak fluxes $S_{\rm{peak}}(3.4\rm{mm})$ of the four sources, mm1
to mm3 and mm6, are listed in Table \ref{6334in_fluxes}. Emission
features below $5\sigma$ are not considered further.  The 3.4\,mm
peak associated with mm1 is the strongest and shows an additional
extension toward the north-western 1.4\,mm source mm5. In contrast to
that, the 1.4\,mm peak mm4, which is associated with cm continuum and
CH$_3$OH class {\sc ii} maser emission, is not detected in our data
above the $5\sigma$ level.

\begin{figure}[htb]
\includegraphics[angle=-90,width=8.8cm]{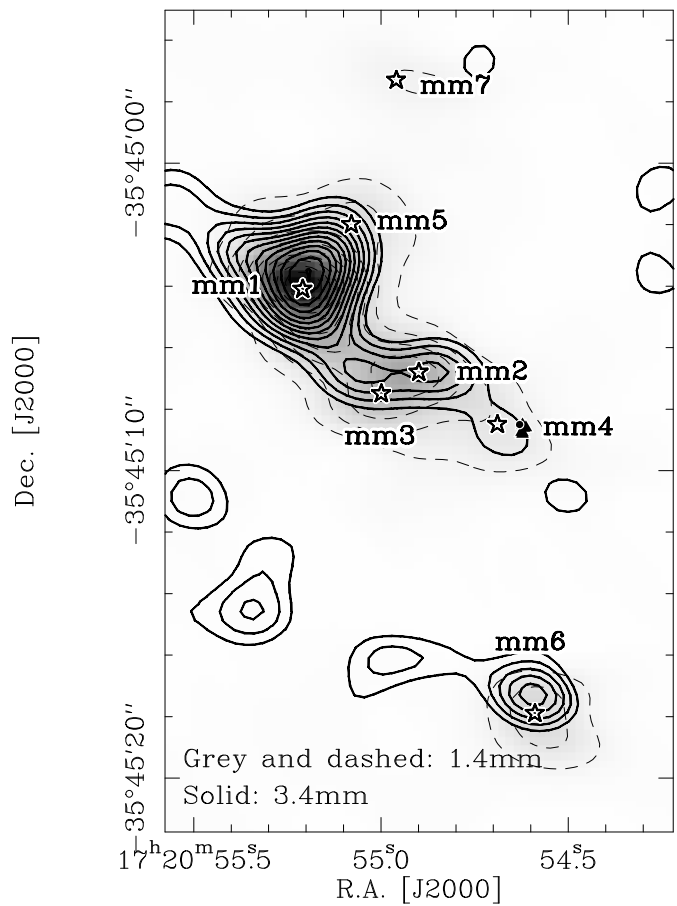}
\caption{Overlay of the SMA 1.4\,mm continuum map (grey-scale with
  dashed contours) with the ATCA 3.4\,mm map (solid contours) at the
  same angular resolution of $2.6''\times 1.8''$. The 1.4\,mm data are
  contoured from 10 to 90\% of the peak emission
  (1077\,mJy\,beam$^{-1}$). The 3.4\,mm emission is contoured in
  $2\sigma$ steps ($2\sigma(3.4\rm{mm})\sim 5.2$\,mJy\,beam$^{-1}$.
  The synthesized beam of both images is the same with $2.6''\times
  1.8''$ (position angle of 84 degrees from north).}
\label{1mm3mm}
\end{figure}

For a better comparison of the 3.4\,mm fluxes with the previously
observed 1.4\,mm observations, we re-imaged the 1.4\,mm data with
exactly the same synthesized beam of $2.6''\times 1.8''$ (position
angle of 84 degrees from north). Figure \ref{1mm3mm} presents an
overlay of the SMA 1.4\,mm data with the ATCA 3.4\,mm data at this
same spatial resolution. While mm1 and mm6 are clearly separated in
both wavelength bands, it is interesting to note that mm2 and mm3
merge in the 1.4\,mm image at the reduced lower spatial resolution.
The corresponding 1.4\,mm peak fluxes are listed in Table
\ref{6334in_fluxes}. For the well separated sources mm1 and mm6 we can
estimate the spectral indices $\alpha$ between both bands now, the
derived values are 3.7 and 3.1, respectively (Table
\ref{6334in_fluxes}). 

Recently \citet{rodriguez2007} showed that the cm emission from
NGC6334I(N) is caused by free-free emission, whereas shortward of
7\,mm wavelength the spectral energy distribution is dominated by dust
continuum emission.  In the Rayleigh-Jeans limit, the flux $S$ scales
with $S\propto\nu^{2+\beta}$ where $\beta$ is the dust opacity index.
With the measured spectral index $\alpha$, we have dust opacity
indices of 1.7 and 1.1 for mm1 and mm6, respectively. While the
$\beta$ value of mm1 is consistent with often observed values between
1.5 and 2, it is lower for mm6.  Although the synthesized beams of
both maps are the same, the uv-coverage was not during the
observations. Furthermore, the continuum maps show a more peaked
morphology toward mm1 than toward mm6. Thus, it is possible that mm6
has more extended structure and that this may be sampled better by the
ATCA observations, possibly accounting for the observed lower values
of $\alpha$ and $\beta$ toward mm6. However, it is also feasible that
physical reasons are responsible for this effect because a decreasing
$\beta$ can be attributed to grain growth in circumstellar disks as
well (e.g., \citealt{beckwith1990}).  Nevertheless, in this scenario,
it appears surprising that mm1, which likely drives an outflow and
hence probably contains an accretion disk, has a value of $\beta$ close to the
interstellar medium values, whereas the source mm6, that does not
exhibit clear signs of ongoing star formation, shows a lower value.
Low $\alpha$ values could in principle also be caused by high optical
depth or an early high-frequency turnover of the Planck-function
caused by the low temperatures.  However, the Planck-turnover of even
a cold core starts changing the slope of the Rayleigh-Jeans part of
the spectral energy distribution usually only above 350\,GHz (e.g.,
\citealt{beuther2007c}), and high optical depth compared to mm1
appears unlikely as well.  Therefore, spatial filtering as well as
physical effects could produce the low $\alpha$ toward mm6, but we
cannot set better constraints here.

It should be noted that the spectral indices we find between 3.4 and
1.3\,mm are larger than those derived previously by
\citet{rodriguez2007} based on VLA 7\,mm data and the 1.3\,mm data by
\citet{hunter2006}. They find $\alpha$ of $\sim$2.4 for both sources
mm1 and mm6. A potential explanation for this discrepancy can arise
from different dust components traced by the various arrays. While the
VLA 7\,mm observations could be dominated by a compact dusty
disk-component that remains optically thick at shorter wavelengths,
the 3.4 and 1.4\,mm data are likely dominated by the larger-scale
optically thin envelope. This envelope with spectral indices between
3.1 and 3.7 (for mm1 and mm6) is weak and below the detection limit at
7\,mm, which can cause the non-detection of any large-scale emission
in the VLA data. A general word of caution should be added that
comparing fluxes from different interferometers may be unreliable
because spatial filtering usually affects the fluxes of most
measurements. Reducing datasets with similar uv-coverage and the same
synthesized beam, as we have done here for the 3.4 and 1.4\,mm data,
helps to minimize this problem.

\begin{table}[htb]
\caption{Millimeter continuum parameters for NGC6334I(N)}
\begin{tabular}{lrrr}
\hline \hline
Source & $S_{\rm{peak}}(3.4\rm{mm})$ & $S_{\rm{peak}}(1.4\rm{mm})$ &$\alpha$ \\
       & [mJy/beam]                  & [mJy/beam] & \\
\hline
mm1 & 36.6 & 1077    & 3.7 \\
mm2 & 14.2 & 524$^a$ & \\
mm3 & 14.2 & 524$^a$ & \\
mm6 & 17.1 & 285     & 3.1 \\
\hline \hline
\end{tabular}
~\\
\footnotesize{$^a$ The sources mm2 and mm3 are merged in the 1.4\,mm dataset.} 
\label{6334in_fluxes}
\end{table}

\subsubsection{Dense gas traced by CH$_3$CN}

The dense gas traced via the CH$_3$CN$(5_K-4_K)$ $K$-ladder exhibits
two prominent peaks associated with the two main mm continuum sources
mm1 and mm2 (Fig.~\ref{ngc6334in}). There is an additional tentative
$3\sigma$ CH$_3$CN peak toward the mm peak mm4 associated with the cm
and CH$_3$OH maser position, however, we refrain from further
interpretation of this because the map shows negative features due to
the incomplete uv-sampling and hence poor deconvolution on a
comparable level. The other 1.4 and 3.4\,mm continuum sources are not
detected in the CH$_3$CN emission. This indicates that the average
densities and/or abundances at those other mm positions are likely
lower than the critical densities of the CH$_3$CN$(5_K-4_K)$ lines of
a few times $10^5$\,cm$^{-3}$ (Table \ref{lines}). Based on this, one
can speculate that except for the sources mm1, mm2 and mm4, all other
mm continuum peaks may still be in a pre-stellar phase prior to active
star formation.

\begin{figure}[htb]
\includegraphics[angle=-90,width=8.8cm]{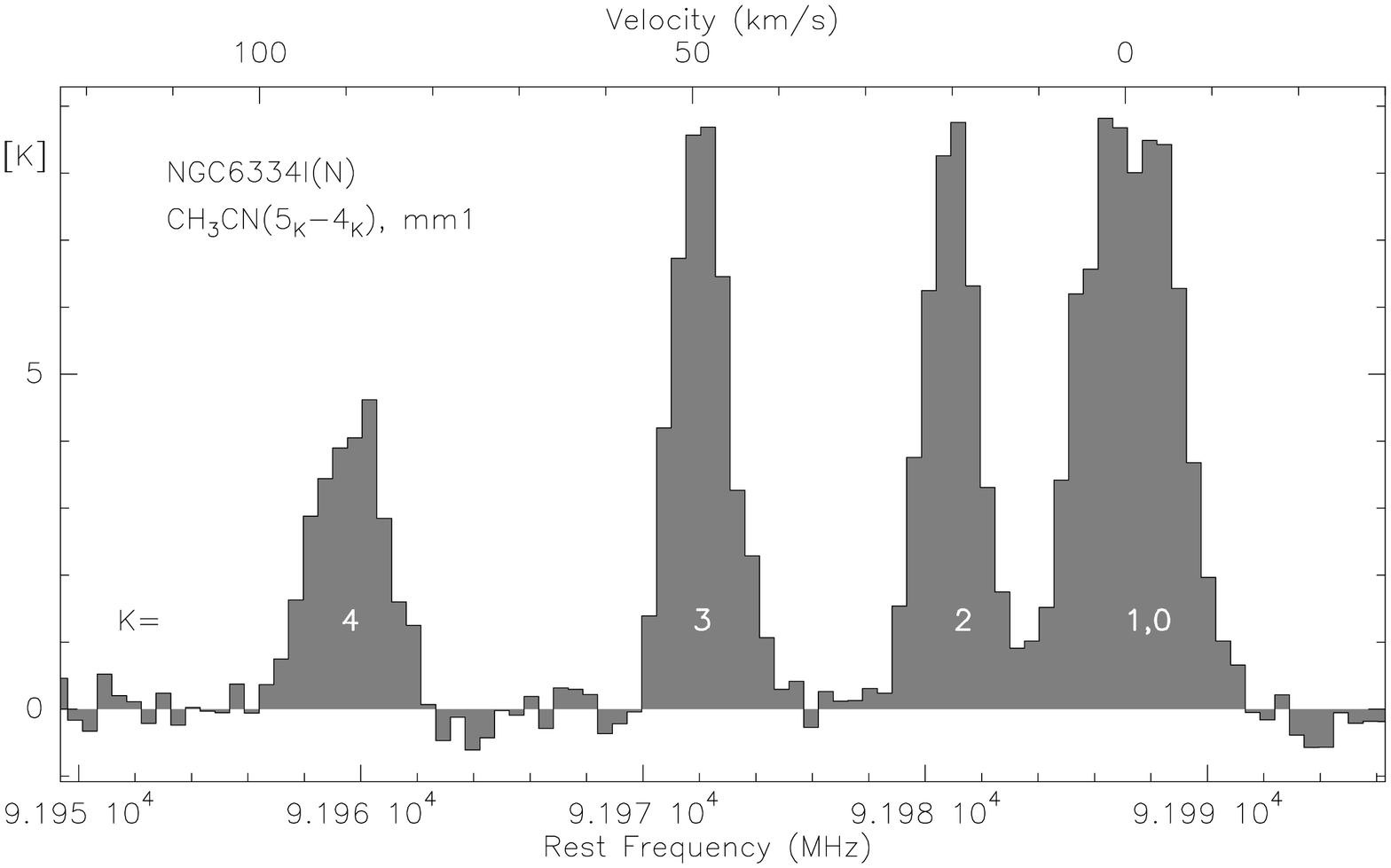}\\
\includegraphics[angle=-90,width=8.8cm]{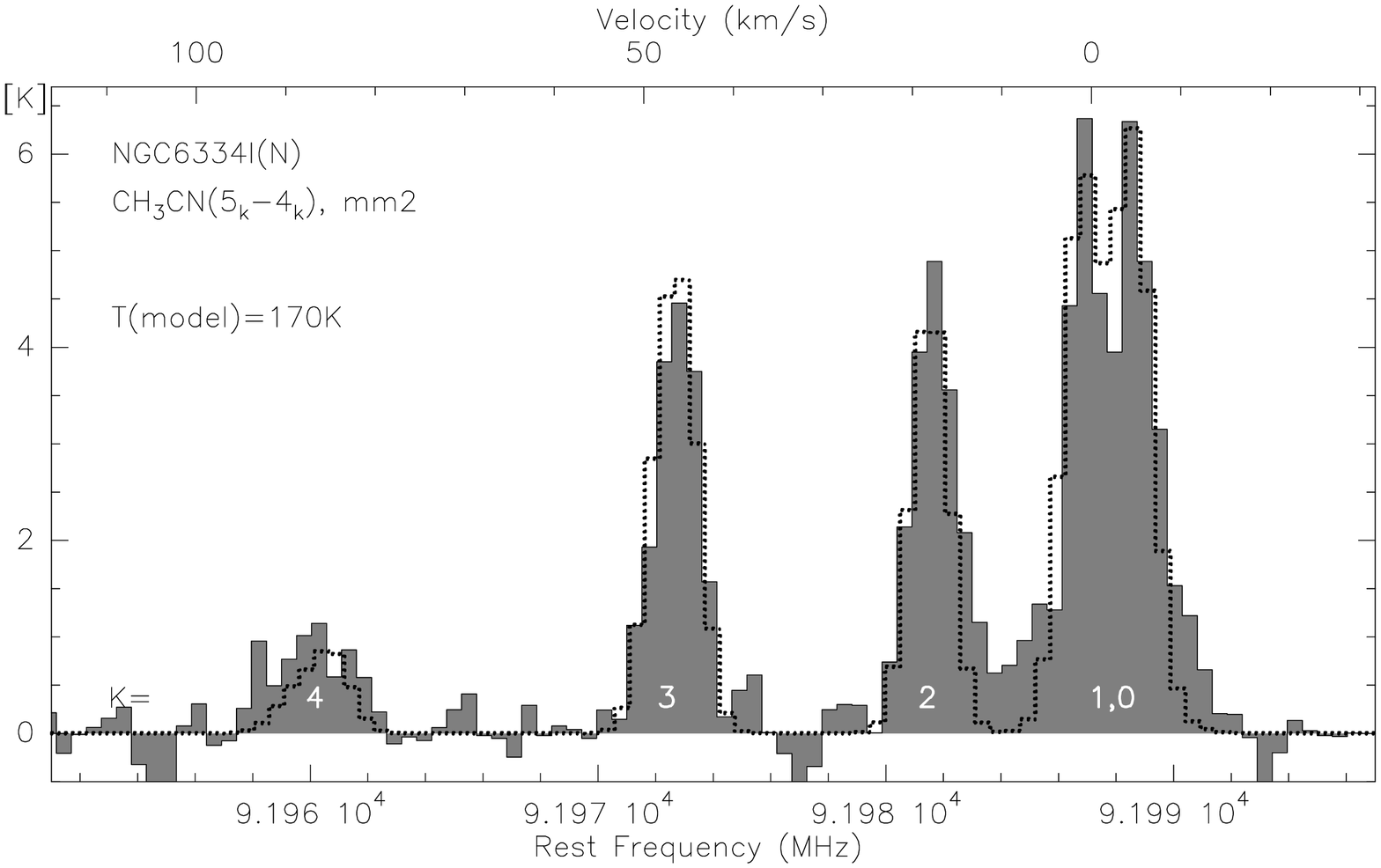}
\caption{CH$_3$CN$(5_K-4_K)$ spectra ($K=0...4$) extracted toward the
  two CH$_3$CN peak positions in NGC6334I(N) shown in Figure
  \ref{ngc6334in}. The dotted line in the lower panel shows a model
  spectrum created with XCLASS at a temperature of 170\,K.}
\label{ngc6334in_ch3cn}
\end{figure}

Figure \ref{ngc6334in_ch3cn} presents the whole CH$_3$CN $K$-ladder
spectrum extracted from the data-cube toward the two peak positions,
and we again detect all $K$-components in both sources. The spectrum
toward mm1 suffers from very high opacity problems as already outlined
for NGC6334I (\S \ref{6334i_ch3cn}), prohibiting any temperature
estimate. However, for mm2 the situation is slightly different and
temperature estimates become feasible. Again using the XCLASS
software, we produced model spectra, and the CH$_3$CN$(5_K-4_K)$
spectrum can be fitted with temperatures in the regime of $170\pm
50$\,K. Such high temperatures additionally confirm the hot core-like
nature of these sub-sources within the overall still relatively cold
region NGC6334I(N) (e.g., \citealt{gezari1982}).

\begin{figure}[htb]
\includegraphics[angle=-90,width=8.8cm]{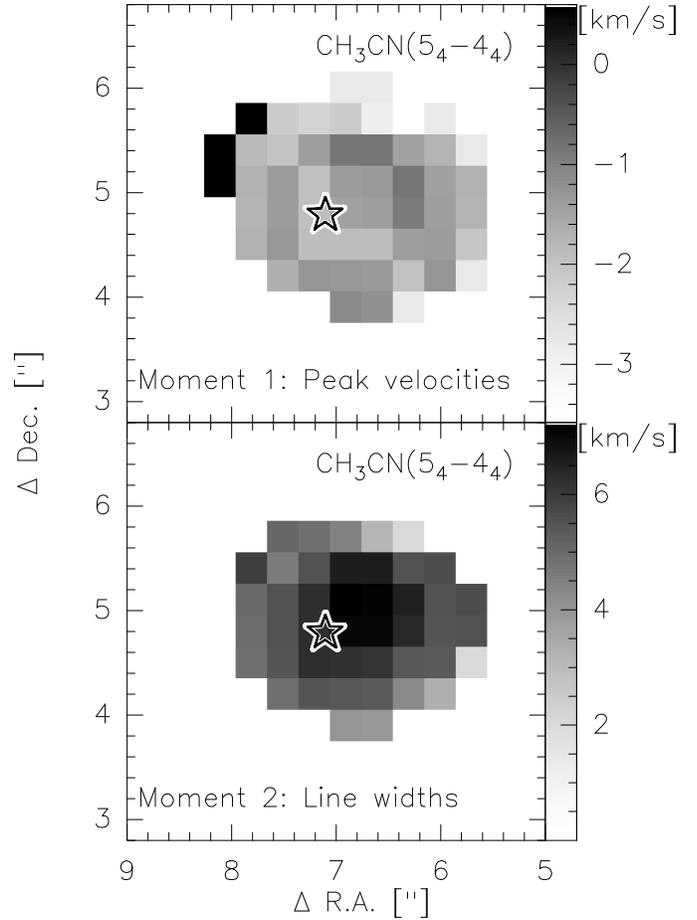}
\caption{Moment maps of CH$_3$CN$(5_K-4_K)$ toward NGC6334I(N) mm1.
  The top panel shows the 1st moment (peak velocities) and the bottom
  panel the 2nd moment (line widths). The star marks the positions of
  the main mm continuum sources from \citet{hunter2006}.}
\label{mom_ch3cn_6334in}
\end{figure}

Since previous NH$_3$(6,6) observations showed a double-horn spectral
profile toward mm1 \citep{beuther2007b}, which could potentially be
produced by an underlying accretion disk, we searched the high-$K$
CH$_3$CN lines for rotation signatures as well. Figure
\ref{mom_ch3cn_6334in} presents the 1st and 2nd moment maps of the
CH$_3$CN$(5_4-4_4)$ line toward mm1. While we see a line-width
increase toward the center, which is indicative of more activity in
the central region (e.g., rotation, outflow, infall), we cannot
identify any obvious velocity gradient. Therefore, the data do not
allow a more detailed analysis of the previously proposed underlying
disk.

\subsubsection{A precessing outflow?}

\begin{figure}[htb]
\includegraphics[angle=-90,width=8.8cm]{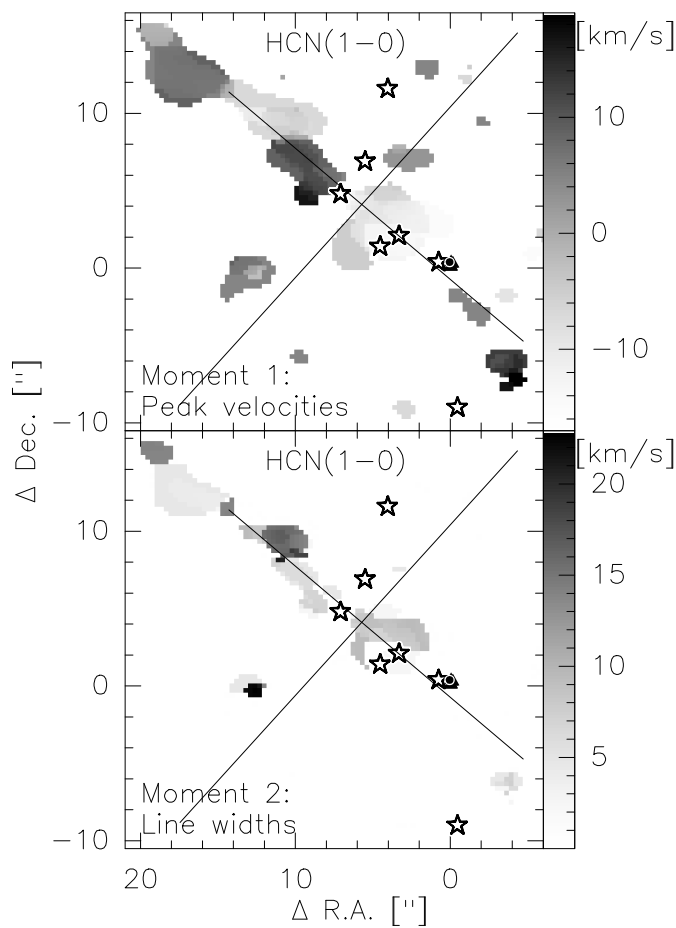}
\caption{Moment maps of HCN(1--0) toward NGC6334I(N). The top panel
  shows the 1st moment (peak velocities) and the bottom panel the 2nd
  moment (line widths). The symbols correspond to the same phenomena
  as in Fig.~\ref{ngc6334in}.}
\label{mom_hcn_6334in}
\end{figure}

The HCN(1--0) emission suffers strongly from the missing short
spacings as indicated by all the negative features seen in Figure
\ref{ngc6334in}. The morphology of the remaining HCN emission
structures nevertheless indicates that most of it is associated with
the north-east south-west molecular outflow than with the
perpendicular one. Figure \ref{mom_hcn_6334in} shows the 1st and 2nd
moment maps of the HCN emission, and the peak velocity and line-width
distributions are a bit peculiar. The most red- and blue-shifted
emission is centered around mm1 indicating that this is likely the
driver of the north-east/south-west outflow.  However, going further
to the north-east along the major outflow axis we find an inversion of
the outflow velocities from red-shifted close to the protostar to
blue-shifted at offsets around ($12''/10''$) from the phase center and
then again red-shifted emission further outward at offsets around
($17''/13''$). Regarding the line-width distribution, the 2nd moment
map in Figure \ref{mom_hcn_6334in} shows that the red-shifted parts of
the north-eastern outflow wing have systematically less broad lines
than the blue-shifted part. This whole behavior is less pronounced in
the south-western outflow direction.

While the line-width differences are hard to explain, the change
between blue-and red-shifted outflow emission is consistent with a
picture in which a mean molecular outflow axis is close to the plane
of the sky, and where the outflow precesses around that axis,
producing in some region the receding red-shifted and in other regions
the approaching blue-shifted emission. This precession picture is
further supported by the spatial distribution of the HCN emission
which resembles a reverse S-shaped morphology in the 1st moment map
(Fig.~\ref{mom_hcn_6334in}).

The 2nd south-east north-west outflow is so far only observed at lower
spatial resolution by \citet{megeath1999}, who did not identify a
driving source for it. While it could emanate from any other source,
e.g. mm2, it is also possible that it originates from the vicinity of
mm1. In the framework of precessing jets, a likely reason to cause the
precession is the existence of multiple embedded objects (e.g.,
\citealt{fendt1998}) which hence could also produce the quadrupolar
outflow morphology (e.g., \citealt{gueth2001}).

\section{Conclusions and Summary}

Our 3.4\,mm continuum and CH$_3$CN/HCN spectral line study of the
massive twin cores NGC6334I and I(N) reveals many new insights into
that intriguing pair of massive star-forming regions. Both sets of
spectral lines as well as the continuum emission are clearly detected
toward both targets. While the continuum emission in NGC6334I mainly
follows the UCH{\sc ii} regions and the strongest protostellar 1.4\,mm
peak is detected only at a $\sim 3\sigma$ level, in NGC6334I(N), the
3.4\,mm continuum emission traces four of the previously identified
protostellar or pre-stellar condensations.

In both regions, the whole CH$_3$CN$(5_K-4_K)$ $K$-ladder from $K=0$
to 4 is detected toward the strongest protostellar condensations.
While the emission is in most cases so optically thick that
temperature estimates are prohibited, toward the secondary mm peak in
NGC6334I(N) we can estimate a temperature of $170\pm50$\,K.  Toward
all four detected CH$_3$CN emission sources, we find a correlation
between increasing line-width and increasing excitation temperature of
the $K$ components. Since increasing excitation temperatures are
expected closer to the protostars, this implies more internal motions,
e.g., outflow, infall or rotation, the closer one gets to the central
protostar. Similar signatures are observed in the CH$_3$CN 2nd moment
maps.

To investigate potential rotation, we produced 1st moment maps toward
all CH$_3$CN peaks, and fitted the channel peak positions toward mm1
in NGC6334I to effectively increase the spatial resolution. We
identify a velocity gradient toward mm1 in NGC6334I that is oriented
approximately perpendicular to the known large-scale outflow. This may
be interpreted as a signature of a rotating structure, maybe
associated with a massive accretion disk. While early rotation-disk
claims for that region were on scales of the molecular outflow
\citep{jackson1988}, we are now reaching spatial scales of the order a
few hundred AU, much more reasonable for accretion disks (e.g.,
\citealt{yorke2002,krumholz2006b}). However, we have to stress that
the rotation signatures are not conclusive yet because, e.g., an
unresolved double-source could produce similar signatures. Further
investigation at higher angular resolution are required to resolve
this issue. While higher angular resolution images with the ATCA at
3\,mm wavelengths are difficult because of decreasing phase stability
with increasing baseline length, one may tackle that problem in some
highly excited NH$_3$ lines in the 12\,mm band. Furthermore, ALMA will
allow the investigation of this source in much greater depth. Toward the
previously suggested disk candidate in NGC6334I(N) we cannot identify
a similar velocity gradient.

In contrast to conventional wisdom that HCN traces the dense gas
cores, we find it most prominently in the molecular outflows of both
massive star-forming regions. The velocity structure of the outflow in
NGC6334I is relatively normal and follows the well-known Hubble-law
for molecular outflows. In addition to that, we find the broadest HCN
line-width toward the main mm continuum peak mm1. In contrast to the
previously found elongation of NH$_3$(6,6) maser emission indicating
that mm2 could drive the molecular outflow, these data suggest that
mm1 harbors the outflow driving source. However, it is also possible
that there are two molecular outflows with different properties that
are preferentially detected in different tracers (HCN in this work and
CO for the previous larger-scale outflow detection at a slightly
different position-angle).  The velocity structure of the NGC6334I(N)
outflow is more peculiar.  There we find a change between blue- and
red-shifted outflow emission on one side of the outflow.  Taking into
account the additionally bended morphology of that outflow, a possible
explanation for this velocity structure is a precessing outflow close
to the plane of the sky.

Furthermore, HCN exhibits a broad absorption feature with a line-width
of $\sim 15.4$\,km\,s$^{-1}$ toward the UCH{\sc ii} region in
NGC6334I. Comparing the line-width of the previously observed
absorption features of NH$_3$ and CH$_3$OH, which are of the order
2\,km\,s$^{-1}$, with the HCN line-width as well as the line-width
observed in the ionized gas of $\sim$32\,km\,s$^{-1}$, two
explanations are possible to explain the different line-widths. The
velocity gradient identified in the ionized gas indicates that it may
be influenced by the molecular outflow close by. If the opacity of HCN
were very large it traced only the outer gas layers around the UCH{\sc
  ii} region and could also be influenced by the outflow. On the other
hand, NH$_3$ is optically thin based on the absent absorption in
the hyperfine satellite lines. If HCN were optically thin as well,
another possible explanation would be based on the different critical
densities of the various molecules: In the picture of an expanding
UCH{\sc ii} region, the densities close to the UCH{\sc ii} region
surface should be higher than those further outside.  Hence HCN may
trace the gas closer to the expanding UCH{\sc ii} region and is then
much stronger affected by the expansion process that the lower-density
regions further out that are traced by NH$_3$ and CH$_3$OH. To
differentiate between both models, observations of a rarer HCN
isotopologues are required to derive its optical depth.

\begin{acknowledgements} 
  We like to thank Peter Schilke for providing the XCLASS software to
  model the CH$_3$CN spectra. Furthermore, we appreciate the careful
  referee's report which helped improving the paper. H.B.~acknowledges
  financial support by the Emmy-Noether-Program of the Deutsche
  Forschungsgemeinschaft (DFG, grant BE2578).
\end{acknowledgements}


\end{document}